\documentclass[aip,amsmath,amssymb,reprint]{revtex4-1}
\usepackage{epsfig}                                                                 
\usepackage{dcolumn}                                                                
\usepackage{bm}                                                                     
\let\boldsymbol\pmb
\usepackage[pdfstartview=FitH,bookmarksopen=true,bookmarksopenlevel=1]{hyperref}    
\begin{document}
\title[The dynamics of liquid films, as described by the diffuse-interface model]%
      {The dynamics of liquid films, as described by the diffuse-interface model}
\author{E. S. Benilov}
 \email[Email address: ]{Eugene.Benilov@ul.ie}
 \homepage[\newline\hspace*{1.4mm} Homepage: ]{https://staff.ul.ie/eugenebenilov/}
 \affiliation{Department of Mathematics and Statistics, University of Limerick,\\Limerick V94~T9PX, Ireland}

\date{\today}

\begin{abstract}
The dynamics of a thin layer of liquid, between a flat solid substrate and an
infinitely-thick layer of saturated vapor, is examined. The liquid and vapor
are two phases of the same fluid, governed by the diffuse-interface model. The
substrate is maintained at a fixed temperature, but in the bulk of the fluid
the temperature is allowed to vary. The slope $\varepsilon$ of the
liquid/vapor interface is assumed to be small, as is the ratio of its
thickness to that of the film. Three asymptotic regimes are identified,
depending on the vapor-to-liquid density ratio $\rho_{v}/\rho_{l}$. If
$\rho_{v}/\rho_{l}\sim1$ (which implies that the temperature is comparable,
but not necessarily close, to the critical value), the evolution of the
interface is driven by the vertical flow due to liquid/vapor phase transition,
with the horizontal flow being negligible. In the limit $\rho_{v}/\rho
_{l}\rightarrow0$, it is the other way around, and there exists an
intermediate regime, $\rho_{v}/\rho_{l}\sim\varepsilon^{4/3}$, where the two
effects are of the same order. Only the $\rho_{v}/\rho_{l}\rightarrow0$ limit
is mathematically similar to the case of incompressible (Navier--Stokes)
liquids, whereas the asymptotic equations governing the other two regimes are
of different types.

\end{abstract}
\maketitle

\section{Introduction}

The diffuse-interface model (DIM) originates from the idea of van der Waals
\cite{Vanderwaals93} and Korteweg \cite{Korteweg01} that intermolecular
attraction in fluids can be modeled by relating it to macroscopic variations
of the fluid density. In recent times, this approach was incorporated into
hydrodynamics: more comprehensive models have been developed for
multi-component fluids with variable temperature
\cite{AndersonMcFaddenWheeler98,ThieleMadrugaFrastia07} -- and simpler ones,
for \emph{single}-component \emph{isothermal} fluids\cite{PismenPomeau00}) or
single-component isothermal and \emph{incompressible} fluids
\cite{JasnowVinals96,Jacqmin99,DingSpelt07,MadrugaThiele09} (in the last case,
the van der Waals force does not depend on the (constant) density, but on a
certain \textquotedblleft order parameter\textquotedblright\ satisfying the
Cahn--Hilliard equation).

Various versions of the DIM have been used in applications, such as
nucleation, growth, and collapse of vapor bubbles
\cite{MagalettiMarinoCasciola15,MagalettiGalloMarinoCasciola16,GalloMagalettiCasciola18,GalloMagalettiCoccoCasciola20}%
, drops impacting on a solid wall \cite{GelissenVandergeldBaltussenKuerten20},
and contact lines in fluids
\cite{Seppecher96,DingSpelt07,YueZhouFeng10,YueFeng11,SibleyNoldSavvaKalliadasis13a,SibleyNoldSavvaKalliadasis13b,SibleyNoldSavvaKalliadasis14,KusumaatmajaHemingwayFielding16,FakhariBolster17,BorciaBorciaBestehornVarlamovaHoefnerReif19}%
.

When studying contact lines, a boundary condition describing the interaction
of the fluid and substrate is needed. Two version of such have been suggested:
one involving the near-substrate density \cite{Seppecher96} and its normal
derivative, and another prescribing just the density \cite{PismenPomeau00}.
The former is based on minimisation of the wall free energy
\cite{PismenPomeau00}, whereas the latter can be obtained through an
asymptotic expansion of the non-local representation of the van der Waals
force \cite{Benilov20a}. In the present paper, the latter (simpler) boundary
condition is used.

The DIM has been also adapted for the case of liquid films, where the liquid
phase is confined to a thin layer bounded by a liquid/vapor interface and a
solid substrate. Assuming that the flow is isothermal and the saturated-vapor
density $\rho_{v}$ is much smaller that the liquid density $\rho_{l}$, Pismen
and Pomeau \cite{PismenPomeau00} derived an asymptotic version of the DIM
similar to the thin-film approximation of the Navier--Stokes equations for
incompressible fluids.

It has been argued, however, that in some, if not most, common fluids
including water, liquid/vapor interfaces are \emph{not} isothermal. Using a
non-isothermal version of the DIM, Refs. \cite{Benilov20a,Benilov20b}
estimated the density and pressure change near the interface and showed that
the resulting temperature change is order-one. It is unclear, however, whether
this conclusion affects liquid films, as a \emph{thin} liquid layer can behave
differently from the general case -- especially, if the substrate is
maintained at a fixed temperature, acting as a thermostat for the adjacent fluid.

There are two more omissions in the existing literature on liquid films with a
diffuse interface. Firstly, no thin-film models exist for the regime with
$\rho_{v}\sim\rho_{l}$ observed at medium and high temperatures. Secondly,
no-one has examined the implications for films of a recently-identified
contradiction between the DIM and the Navier--Stokes equations: as shown in
Ref. \cite{Benilov20c}, the former does not admit solutions describing static
two-dimensional sessile drops (also called liquid ridges), whereas the latter
do. A similar comparison between the thin-film asymptotics of the two models
should clarify the nature of the discrepancy, as asymptotic models are much
simpler than the exact ones.

The present paper tackles the above omissions. It is shown that, if $\rho
_{v}\sim\rho_{l}$, the heat released (consumed) due to the fluid compression
(expansion) near the interface makes non-isothermality important, so the
thin-film asymptotics in this case differs from that derived in Ref.
\cite{PismenPomeau00}. In the limit $\rho_{v}/\rho_{l}\rightarrow0$, however,
liquid films are essentially isothermal and the thin-film approximation of the
DIM coincides with that of the Navier--Stokes equations. This implies that
liquid ridges exist in the former model as quasi-static states, i.e., they
evolve, but so slowly that the evolution is indistinguishable from, say, evaporation.

The present paper is structured as follows. In Sect. \ref{Sect. 2}, the
problem is formulated mathematically, and in Sect. \ref{Sect. 3}, the simplest
case of static interfaces is examined. The regimes $\rho_{v}\sim\rho_{l}$ and
$\rho_{v}\ll\rho_{l}$ are examined in Sects. \ref{Sect. 4} and \ref{Sect. 5},
respectively. Since these sections include a lot of cumbersome algebra, a
brief summary of the results, plus their extensions to three-dimensional
flows, are presented in Sect. \ref{Sect. 6}.

\section{Formulation\label{Sect. 2}}

Consider a compressible fluid flow characterized by the density $\rho
(\mathbf{r},t)$, velocity $\mathbf{v}(\mathbf{r},t)$, and temperature
$T(\mathbf{r},t)$, where $\mathbf{r}$ is the position vector and $t$, the
time. Let the pressure $p$ be related to $\rho$ and $T$ by the the van der
Waals equation of state,%
\begin{equation}
p=\frac{RT\rho}{1-b\rho}-a\rho^{2}, \label{2.1}%
\end{equation}
where $R$ is the specific gas constant, and $a$ and $b$ are fluid-specific
constants ($b$ is the reciprocal of the maximum allowable density). Eq.
(\ref{2.1}) was chosen for its simplicity, with all of the results obtained
below being readily extendable to general non-ideal fluids.

The diffuse-interface model in application to compressible Newtonian fluid is
\cite{AndersonMcFaddenWheeler98}%
\begin{equation}
\frac{\partial\rho}{\partial t}+\boldsymbol{\boldsymbol{\nabla}}\cdot\left(
\rho\mathbf{v}\right)  =0, \label{2.2}%
\end{equation}%
\begin{equation}
\rho\left[  \frac{\partial\mathbf{v}}{\partial t}+\left(  \mathbf{v}%
\cdot\boldsymbol{\boldsymbol{\nabla}}\right)  \mathbf{v}\right]
+\boldsymbol{\boldsymbol{\nabla}}p-\boldsymbol{\boldsymbol{\nabla}}%
\cdot\boldsymbol{\Pi}=K\rho\boldsymbol{\boldsymbol{\nabla}}\nabla^{2}\rho,
\label{2.3}%
\end{equation}%
\begin{multline}
\rho c_{V}\left(  \frac{\partial T}{\partial t}+\mathbf{v}\cdot
\boldsymbol{\boldsymbol{\nabla}}T\right)  +\left[  \mathbf{I}\left(
p+a\rho^{2}\right)  -\boldsymbol{\Pi}\right]  :\boldsymbol{\boldsymbol{\nabla
}}\mathbf{v}\\
-\boldsymbol{\boldsymbol{\nabla}}\cdot\left(  \kappa
\boldsymbol{\boldsymbol{\nabla}}T\right)  =0, \label{2.4}%
\end{multline}
where $\mathbf{I}$ is the identity matrix,%
\begin{equation}
\boldsymbol{\Pi}=\mu_{s}\left[  \boldsymbol{\boldsymbol{\nabla}}%
\mathbf{v}+\left(  \boldsymbol{\boldsymbol{\nabla}}\mathbf{v}\right)
^{T}-\frac{2}{3}\mathbf{I}\left(  \boldsymbol{\boldsymbol{\nabla}}%
\cdot\mathbf{v}\right)  \right]  +\mu_{b}\,\mathbf{I}\left(
\boldsymbol{\boldsymbol{\nabla}}\cdot\mathbf{v}\right)  \label{2.5}%
\end{equation}
is the viscous stress tensor, $K$ is the so-called Korteweg parameter,
$\mu_{s}$ ($\mu_{b}$) is the shear (bulk) viscosity, $c_{V}$ is the specific
heat capacity, and $\kappa$, the thermal conductivity. Note that, generally,
$\mu_{s}$, $\mu_{b}$, $c_{V}$, and $\kappa$ depend on $\rho$ and $T$, whereas
$K$ is a constant.

In what follows, two-dimensional flows will be mainly explored, so
$\mathbf{r}=\left[  x,z\right]  $ and $\mathbf{v}=\left[  u,w\right]  $ where
$x$ and $u$ are the horizontal components of the corresponding vectors, and
$z$ and $w$ are their vertical components. The three-dimensional extensions of
the results obtained will be presented without derivation in Sect.
\ref{Sect. 6}.

Assume that the fluid is bounded below by a solid substrate located at $z=0$,
so the flow is constrained by%
\begin{equation}
\mathbf{v}=\mathbf{0}\qquad\text{at}\qquad z=0,\label{2.6}%
\end{equation}%
\begin{equation}
T=T_{0}\qquad\text{at}\qquad z=0,\label{2.7}%
\end{equation}%
\begin{equation}
\rho=\rho_{0}\qquad\text{at}\qquad z=0.\label{2.8}%
\end{equation}
(\ref{2.6}) is the no-flow boundary condition, (\ref{2.7}) implies that the
substrate is maintained at a fixed temperature $T_{0}$, and the near-wall
density $\rho_{0}$ in (\ref{2.8}) is a phenomenological parameter (in the
diffuse-interface model \cite{PismenPomeau00,Benilov20a}, it is assumed to be
known). Note that the parameter $\rho_{0}$ is specific to the fluid--substrate
combination under consideration and is uniquely related to the contact angle.

Given a suitable initial condition, the boundary-value problem (\ref{2.1}%
)-(\ref{2.8}) determines the unknowns $\rho(\mathbf{r},t)$, $\mathbf{v}%
(\mathbf{r},t)$, and $T(\mathbf{r},t)$.

\section{Static films\label{Sect. 3}}

Before examining the evolution of liquid films, it is instructive to briefly
review the properties of static films.

Letting $\mathbf{v}=\mathbf{0}$ and $\partial\rho/\partial t=0$, and taking
into account that only isothermal films can be static (hence, $T=T_{0}$), one
can reduce Eqs. (\ref{2.1})-(\ref{2.5}) to a single equation%
\begin{equation}
RT_{0}\left(  \ln\frac{b\rho}{1-b\rho}+\frac{1}{1-b\rho}\right)
-2a\rho-K\nabla^{2}\rho=G,\label{3.1}%
\end{equation}
where $G$ is a constant of integration (physically, the free-energy density).
Once Eq. (\ref{3.1}) is complemented with boundary conditions, one can
determine $G$ together with the solution $\rho$.

The one- and two-dimensional solutions of Eq. (\ref{3.1}) will be examined in
Sects. \ref{Sect. 3.1} and \ref{Sect. 3.2}, respectively.

\subsection{Films with flat interfaces\label{Sect. 3.1}}

Let $\rho$ be independent of $x$, so that $\rho(z)$ describes a flat interface
parallel to the substrate. The following nondimensional variables will be
used:%
\begin{equation}
\rho_{nd}=b\rho,\qquad z_{nd}=\frac{z}{z_{0}}, \label{3.2}%
\end{equation}
where%
\begin{equation}
z_{0}=\sqrt{\frac{K}{a}} \label{3.3}%
\end{equation}
is, physically, the characteristic thickness of liquid/vapor interfaces.
Estimates of $z_{0}$ for specific applications presented in Refs.
\cite{MagalettiGalloMarinoCasciola16,Benilov20a} show that $z_{0}$ is on a
nanometer scale; hereinafter it will be referred to as \textquotedblleft
microscopic\textquotedblright.

\begin{figure}
\includegraphics[width=\columnwidth]{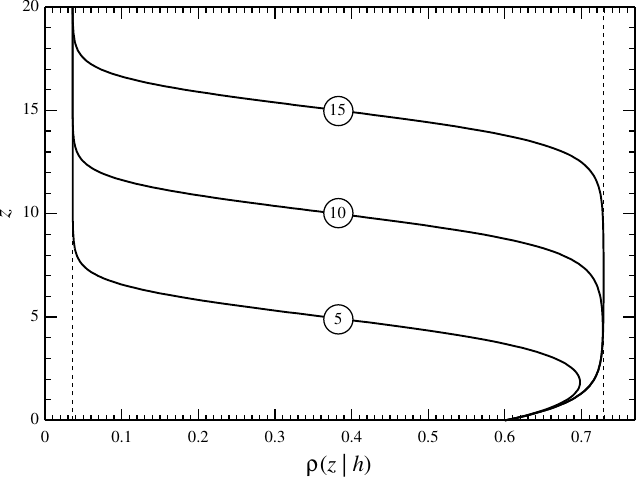}
\caption{The function $\rho(z|h)$ [determined by (\ref{3.5}), (\ref{2.8}), (\ref{3.7})-(\ref{3.8})] for $\tau=0.2$ and $\rho_{0}=0.6$. The curves are labelled with the corresponding values of $h$. The dotted lines show $\rho_{v}$ and $\rho_{l}$.}
\label{fig1}
\end{figure}

It is convenient to also introduce the nondimensional analogues of the
parameters $\rho_{0}$ and $G$,%
\begin{equation}
\left(  \rho_{0}\right)  _{nd}=b\rho_{0},\qquad G_{nd}=\frac{b}{a}%
G.\label{3.4}%
\end{equation}
In nondimensional form, Eq. (\ref{3.1}) is (the subscript $_{nd}$ omitted)%
\begin{equation}
\tau\left(  \ln\frac{\rho}{1-\rho}+\frac{1}{1-\rho}\right)  -2\rho
-\frac{\mathrm{d}^{2}\rho}{\mathrm{d}z^{2}}=G,\label{3.5}%
\end{equation}
where the first two terms on the left-hand side represent the nondimensional
free-energy density of the van der Waals fluid, and%
\begin{equation}
\tau=\frac{RT_{0}b}{a}\label{3.6}%
\end{equation}
is the nondimensional temperature. The nondimensional version of the boundary
condition (\ref{2.8}) will not be presented as it looks exactly as its
dimensional counterpart. One should also impose the requirement of zero
Korteweg stress at infinity,%
\begin{equation}
\frac{\mathrm{d}\rho}{\mathrm{d}z}\rightarrow0\qquad\text{as}\qquad
z\rightarrow\infty.\label{3.7}%
\end{equation}
Due to the presence of the undetermined constant $G$, Eq. (\ref{3.5}) and the
boundary conditions (\ref{2.8}), (\ref{3.7}) do not fully determine the
solution. The most convenient way to fix $\rho(z)$ consists in prescribing the
height $h$ of the interface through the requirement%
\begin{equation}
\rho(h)=\frac{\max\left\{  \rho(z)\right\}  +\min\left\{  \rho(z)\right\}
}{2}.\label{3.8}%
\end{equation}
In what follows, the solution of the boundary-value problem (\ref{3.5}),
(\ref{2.8}), (\ref{3.7})-(\ref{3.8}) will be denoted by $\rho(z|h)$. Several
examples of $\rho(z|h)$ with increasing $h$ are shown in Fig. \ref{fig1}.

In this work, the following properties of $\rho(z|h)$ will be
needed:\smallskip

(1) Since $z$ was nondimensionalized on a microscopic scale, \emph{macro}%
scopic films correspond to $h\gg1$, making this limit important for
applications -- both industrial (e.g., paint or polymer coating) and natural
(e.g., rainwater flowing down a rockface).

For large $h$, the interface is located far from the substrate, so the
interfacial profile is similar to that in an unbounded fluid. Mathematically,
this means%
\begin{equation}
\rho(z|h)\rightarrow\bar{\rho}(z-h)\qquad\text{as}\qquad z,h\rightarrow
\infty,\label{3.9}%
\end{equation}
where $\bar{\rho}(z)$ satisfies the same equation as $\rho(z,h)$ and the
open-space boundary conditions,%
\begin{equation}
\tau\left(  \ln\frac{\bar{\rho}}{1-\bar{\rho}}+\frac{1}{1-\bar{\rho}}\right)
-2\bar{\rho}-\frac{\mathrm{d}^{2}\bar{\rho}}{\mathrm{d}z^{2}}=G,\label{3.10}%
\end{equation}%
\begin{align}
\bar{\rho}(z) &  \rightarrow\rho_{l}\qquad\,\text{as}\qquad z\rightarrow
-\infty,\label{3.11}\\
\bar{\rho}(z) &  \rightarrow\rho_{v}\qquad\text{as}\qquad z\rightarrow
\infty,\label{3.12}%
\end{align}%
\begin{equation}
\bar{\rho}(0)=\frac{\rho_{l}+\rho_{v}}{2}.\label{3.13}%
\end{equation}
Eqs. (\ref{3.10})-(\ref{3.13}) fix $\bar{\rho}(z)$, as well as $\rho_{v}$ and
$\rho_{l}$ (which represent the nondimensional densities of saturated vapor
and liquid, respectively).

As observed in Ref. \cite{PismenPomeau00}, the influence of the substrate
decays exponentially with the distance, which implies that the asymptotic
formula (\ref{3.9}) is accurate even for moderate (logarithmically large) $h$.
However, even though $\bar{\rho}(z-h)$ approximates $\rho(z|h)$ well in the
interfacial region, $\bar{\rho}(z-h)$ does not generally satisfy the boundary
condition at the substrate. The only exception is the case where $\rho_{0}$ is
close to $\rho_{l}$, which implies that near the substrate, $\rho
(z|h)\approx\bar{\rho}(z-h)+\mathcal{O}(\varepsilon)$, where $\varepsilon
=\rho_{l}-\rho_{0}$. Merging this result with (\ref{3.9}) (which is
exponentially accurate in both $1/h$ and $\varepsilon$), one obtains%
\begin{equation}
\rho(z|h)=\bar{\rho}(z-h)+\mathcal{O}(\varepsilon)\qquad\text{if}\qquad h\gg1,
\label{3.14}%
\end{equation}
which applies to all $z$. Note also that the limit of small $\varepsilon$ is
important as it corresponds to the approximation of small contact angle (more
details are given below).\smallskip

(2) $\rho_{l}$ and $\rho_{v}$ can be computed without calculating $\bar{\rho
}(z)$, through the so-called Maxwell construction. In the low-temperature
limit $\tau\rightarrow0$, it yields (see Appendix \ref{Appendix A})%
\begin{equation}
\rho_{l}=\frac{1+\sqrt{1-4\tau}}{2}+\mathcal{O}(\operatorname{e}^{-1/\tau}),
\label{3.15}%
\end{equation}%
\begin{equation}
\rho_{v}=\frac{1+\sqrt{1-4\tau}}{1-\sqrt{1-4\tau}}\operatorname{e}^{-1/\tau
}+\mathcal{O}(\tau^{-1}\operatorname{e}^{-2/\tau}). \label{3.16}%
\end{equation}
Thus, if $\tau$ is small, $\rho_{v}$ is \emph{exponentially} small.

If $\tau$ increases, $\rho_{v}$ grows and $\rho_{l}$ decays; eventually, they
merge at the critical point $\left(  \rho_{v}\right)  _{cr}=\left(  \rho
_{l}\right)  _{cr}=1/3$, $\tau_{cr}=8/27$. For larger $\tau$, only one phase
exists, so liquid films do \emph{not} exists.

For $\tau\ll1$, one can also obtain an exponentially accurate expression for
the whole solution $\bar{\rho}(z)$, but it is bulky and implicit. In what
follows, an algebraically accurate but explicit expression will be
used,\begin{widetext}%
\begin{equation}
\bar{\rho}(z)=\left\{
\begin{tabular}
[c]{ll}%
$1+\mathcal{O}(\tau\ln\tau)\medskip$ & if$\qquad z\leq-2^{-3/2}\pi,$\\
$\frac{1}{2}\left(  1-\sin2^{1/2}z\right)  +\mathcal{O}(\tau\ln\tau
)\medskip\qquad$ & if$\qquad z\in\left[  -2^{-3/2}\pi,~2^{-3/2}\pi\right]
,$\\
$0+\mathcal{O}(\tau\ln\tau)$ & if$\qquad z\geq2^{-3/2}\pi.$%
\end{tabular}
\ \right.  \label{3.17}%
\end{equation}
\end{widetext}This solution follows from Eq. (\ref{3.10}) with $\tau=0$ and
the boundary conditions (\ref{3.11})-(\ref{3.12}) with $\rho_{l}=1$ and
$\rho_{v}=0$.

The low-temperature limit is important, as $\tau$ is indeed small for many
common liquids. For water at $20^{\circ}\mathrm{C}$, for example, estimates of
$\tau$ vary from $0.064$ to $0.14$ (depending on the equation of state used --
see Refs. \cite{Benilov20a,Benilov20b}).\smallskip

(3) In what follows, the function%
\begin{equation}
\rho^{\prime}(h)=\frac{1}{\rho_{l}-\rho_{0}}\left[  \frac{\mathrm{d}\rho
(z|h)}{\mathrm{d}z}\right]  _{z=0} \label{3.19}%
\end{equation}
plays an important role. It can be readily computed -- see examples shown in
Fig. \ref{fig2}. Evidently, $\rho^{\prime}(h)$ is bounded above, and its
precise upper bound is (see Appendix \ref{Appendix B})
\begin{multline}
\rho^{\prime}(h)<\frac{\sqrt{2}}{\rho_{l}-\rho_{0}}\\
\times\sqrt{\tau\left[  \rho_{0}\ln\dfrac{\rho_{0}\left(  1-\rho_{v}\right)
}{\rho_{v}\left(  1-\rho_{0}\right)  }-\frac{\rho_{0}-\rho_{v}}{1-\rho_{v}%
}\right]  -\left(  \rho_{0}-\rho_{v}\right)  ^{2}}. \label{3.20}%
\end{multline}
As follows from Fig. \ref{fig2}, $\rho^{\prime}(h)$ tends to its maximum as
$h\rightarrow\infty$. What happens in this limit with $\rho(z|h)$ has been
illustrated in Fig. \ref{fig1}.

\begin{figure}
\includegraphics[width=\columnwidth]{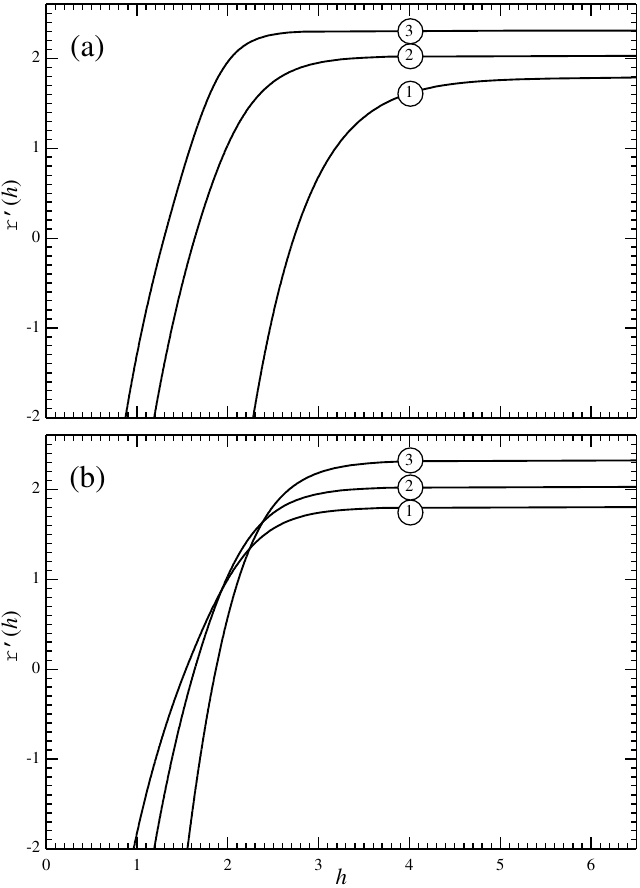}
\caption{The function $\rho^{\prime}(h)$. Panel (a): $\rho_{0}=0.8$, curves 1-3 correspond to $\tau=0.15,~0.1,~0.05$; Panel (b): $\tau=0.1$, curves
1-3 correspond to $\rho_{0}=0.75,~0.8,~0.85$.}
\label{fig2}
\end{figure}

Note that $\rho^{\prime}(h)$ remains order-one in the limit $\rho
_{0}\rightarrow\rho_{l}$. Indeed, letting%
\begin{equation}
\rho_{0}=\rho_{l}-\varepsilon\label{3.21}%
\end{equation}
and expanding estimate (\ref{3.20}) in powers of $\varepsilon$, one can take
into account the Maxwell construction (\ref{A.1})-(\ref{A.2}) and see that the
first two orders of the expansion in $\varepsilon$ vanish, so that
(\ref{3.20}) becomes%
\begin{equation}
\rho^{\prime}(h)<\sqrt{\frac{\tau}{\rho_{l}\left(  1-\rho_{l}\right)  ^{2}}%
-2}+\mathcal{O}(\varepsilon). \label{3.22}%
\end{equation}
The limit $\varepsilon\ll1$ is particularly important, as it corresponds to
the contact angle being small \cite{PismenPomeau00}.

\subsection{Films with slightly curved interfaces\label{Sect. 3.2}}

Consider the full (two-dimensional) equation (\ref{3.1}) and assume that the
interface is curved, but its slope is small. This can only occur if the
contact angle is small -- which, in turn, implies that $\rho_{0}$ is close to,
but still smaller than, the liquid density $\rho_{l}$.

Given scaling (\ref{3.2}) of the vertical coordinate $z$, the scaling of the
horizontal coordinate should be%
\begin{equation}
x_{nd}=\frac{\varepsilon x}{z_{0}},\label{3.23}%
\end{equation}
where $\varepsilon\ll1$ is related to the physical parameters by (\ref{3.21}),
but also playes the role of the slope of the interace. Rewriting Eq.
(\ref{3.1}) in terms of the nondimensional variables (\ref{3.2})-(\ref{3.4})
and (\ref{3.23}), one obtains (the subscript $_{nd}$ omitted)%
\begin{equation}
\tau\left(  \ln\frac{\rho}{1-\rho}+\frac{1}{1-\rho}\right)  -2\rho
-\varepsilon^{2}\frac{\partial^{2}\rho}{\partial x^{2}}-\frac{\partial^{2}%
\rho}{\partial z^{2}}=G.\label{3.24}%
\end{equation}
In addition to the boundary condition (\ref{2.8}) at the wall, a condition is
required as $z\rightarrow\infty$. Assuming that the liquid film is bounded
above by an infinite layer of saturated vapor, require%
\begin{equation}
\rho\rightarrow\rho_{v}\qquad\text{as}\qquad z\rightarrow\infty.\label{3.25}%
\end{equation}
This boundary condition is consistent with Eq. (\ref{3.24}) only if%
\begin{equation}
G=\tau\left(  \ln\frac{\rho_{v}}{1-\rho_{v}}+\frac{1}{1-\rho_{v}}\right)
-2\rho_{v}.\label{3.26}%
\end{equation}
The difference between Eq. (\ref{3.24}) and its one-dimensional counterpart
(\ref{3.5}) is $\mathcal{O}(\varepsilon^{2})$ -- hence, the solution of the
former can be sought using that of the latter,%
\begin{equation}
\rho(x,z)=\rho(z|h)+\mathcal{O}(\varepsilon^{2}),\label{3.27}%
\end{equation}
where $h=h(x)$ is an undetermined function. Physically, solution (\ref{3.27})
describes a liquid film with a slowly changing thickness.

Let $h\gg1$, in which case expressions (\ref{3.27}) and (\ref{3.14}) yield%
\begin{equation}
\rho=\bar{\rho}(z-h)+\mathcal{O}(\varepsilon).\label{3.28}%
\end{equation}
This approximation will be used everywhere in this paper. It applies to films
whose dimensional thickness exceeds the thickness $z_{0}$ of the liquid/vapor
interface given by (\ref{3.3}) -- hence, since $z_{0}$ is of on a nanometer
scale, this assumption is not very restrictive.

There are two ways to determine $h(x)$. Firstly, one can expand the solution
in $\varepsilon$, with the leading order determined by (\ref{3.28}) -- then
try to find the next-to-leading-order solution. The latter is likely to exist
only subject to $h(x)$ satisfying a certain differential equation.

Secondly, one can try to rearrange the exact boundary-value problem in such a
way that all leading-order terms cancel; then substitute the leading-order
solution (\ref{3.28}) in the resulting equation(s). For the static case, the
second approach is only marginally simpler -- but, for evolving films, it is
\emph{much} simpler, and so will be used in both cases.

To eliminate the leading-order terms from Eq. (\ref{3.24}), multiply it by
$\partial\rho/\partial z$, integrate from $z=0$ to $z=\infty$, then take into
account the boundary conditions (\ref{3.25}), (\ref{2.8}) and expression
(\ref{3.26}) for $G$. After straightforward algebra, one obtains%
\begin{equation}
-\varepsilon^{2}\frac{\mathrm{d}}{\mathrm{d}x}\int_{0}^{\infty}\frac
{\partial\rho}{\partial x}\frac{\partial\rho}{\partial z}\mathrm{d}%
z=C-\frac{1}{2}\left[  \left(  \frac{\partial\rho}{\partial z}\right)
_{z=0}\right]  ^{2}, \label{3.29}%
\end{equation}
where%
\begin{multline}
C=\tau\left[  \rho_{0}\left(  \ln\frac{\rho_{0}}{1-\rho_{0}}-\ln\frac{\rho
_{v}}{1-\rho_{v}}\right)  -\frac{\rho_{0}-\rho_{v}}{1-\rho_{v}}\right] \\
-\left(  \rho_{0}-\rho_{v}\right)  ^{2}. \label{3.30}%
\end{multline}
Next, substitute (\ref{3.21}) into (\ref{3.30}), expand it in $\varepsilon$,
take into account the Maxwell construction (\ref{A.1})-(\ref{A.2}), and thus
obtain%
\begin{equation}
C=\varepsilon^{2}\left[  \frac{\tau}{2\rho_{l}\left(  1-\rho_{l}\right)  ^{2}%
}-1\right]  +\mathcal{O}(\varepsilon^{3}). \label{3.31}%
\end{equation}
Observe that, even though the exact expression for $C$ involves $\rho_{v}$,
the approximate one involves $\rho_{l}$ (which occurs due to the use of the
Maxwell construction inter-relating these parameters).

Now, substitute the leading-order solution (\ref{3.28}) into Eq. (\ref{3.29})
and take into account (\ref{3.31}). Omitting small terms, one obtains%
\begin{equation}
\frac{\mathrm{d}}{\mathrm{d}x}\left(  \sigma\frac{\mathrm{d}h}{\mathrm{d}%
x}\right)  =\frac{\tau}{2\rho_{l}\left(  1-\rho_{l}\right)  ^{2}}-1-\frac
{1}{2}\rho^{\prime2}(h), \label{3.32}%
\end{equation}
where the function $\rho^{\prime}(h)$ is defined by (\ref{3.19}) and
(\ref{3.21}), and%
\[
\sigma=\int_{0}^{\infty}\left[  \frac{\partial\bar{\rho}(z-h)}{\partial
z}\right]  ^{2}\mathrm{d}z.
\]
Since $h\gg1$, one can extend the above integral to $-\infty$ (without
altering significantly its value),%
\begin{equation}
\sigma=\int_{-\infty}^{\infty}\left[  \frac{\partial\bar{\rho}(z)}{\partial
z}\right]  ^{2}\mathrm{d}z. \label{3.33}%
\end{equation}
This expression does not depend on $h$ and coincides with the capillary
coefficient (see Ref. \cite{PismenPomeau00}).

Finally, substituting (\ref{3.31}) into (\ref{3.32}) one obtains%
\begin{equation}
2\sigma\frac{\mathrm{d}^{2}h}{\mathrm{d}x^{2}}=\frac{\tau}{\rho_{l}\left(
1-\rho_{l}\right)  ^{2}}-2-\rho^{\prime2}(h). \label{3.34}%
\end{equation}
This equation determines the profile $h(x)$ of a liquid film. Bar notation, it
coincides with Eq. (42) of Ref. \cite{PismenPomeau00}, and they both are
thin-film reductions of the requirement that a steady distribution of density
must have homogeneous chemical potential.

\subsection{Does Eq. (\ref{3.34}) admit ridge solutions?\label{Sect. 3.3}}

The most surprising feature of Eq. (\ref{3.34}) is that it does not admit
solutions describing two-dimensional sessile drops (also called liquid
ridges). This conclusion is highly counter-intuitive, as the Navier--Stokes
equations do admit such solutions. This paradox will be resolved in Sect.
\ref{Sect. 5}.

To prove the nonexistence of ridge solutions, note that the DIM does not allow
the substrate to be completely dry \cite{PismenPomeau00}. Hence, ridge
solutions should involve a \textquotedblleft precursor film\textquotedblright,
i.e.,%
\begin{equation}
h\rightarrow h_{pf}\qquad\text{as}\qquad x\rightarrow\pm\infty, \label{3.35}%
\end{equation}
where $h_{pf}$ is the precursor film's thickness. This boundary condition is
consistent with Eq. (\ref{3.34}) only if $h_{pf}$ satisfies either%
\begin{equation}
\rho^{\prime}(h_{pf})=-\sqrt{\frac{\tau}{\rho_{l}\left(  1-\rho_{l}\right)
^{2}}-2} \label{3.36}%
\end{equation}
or%
\begin{equation}
\rho^{\prime}(h_{pf})=\sqrt{\frac{\tau}{\rho_{l}\left(  1-\rho_{l}\right)
^{2}}-2}. \label{3.37}%
\end{equation}
It can be deduced from the Maxwell construction that%
\[
\frac{\tau}{\rho_{l}\left(  1-\rho_{l}\right)  ^{2}}>2\qquad\text{if}%
\qquad\tau<\frac{8}{27},
\]
hence, Eq. (\ref{3.36}) admits a real solution for $h_{pf}$. Eq. (\ref{3.37}),
on the other hand, does \emph{not} admit real solutions due to inequality
(\ref{3.22}).

The mere fact that there exists only one value of $h$ such that the right-hand
side of Eq. (\ref{3.34}) vanishes disallows the existence of ridge solutions.
Indeed, let the ridge's crest be located at $x=0$, i.e.,%
\[
\frac{\mathrm{d}h}{\mathrm{d}x}=0,\qquad h>h_{pf}\qquad\text{at}\qquad x=0,
\]
and assume that $h(x)$ monotonically grows in $\left(  -\infty,0\right)  $ and
decays in $\left(  0,\infty\right)  $. This implies existence of two
inflection points, where $\mathrm{d}^{2}h/\mathrm{d}x^{2}=0$ and $h>h_{pf}$.
The former condition can only hold if the right-hand side of Eq. (\ref{3.34})
vanishes at the inflection points -- which is, however, impossible since it
vanishes only if $h=h_{pf}$.

If the ridge profile involves oscillations and, thus, several pairs of
inflection points, the same argument applies to the farthest one from the
crest (because $h$ at this inflection point certainly differs from $h_{pf}$).
Overall, the conclusion about the nonexistence of \emph{thin} ridges agrees
with a similar result proved in Ref. \cite{Benilov20c} for \emph{arbitrary}
ridges. The physical implications of the non-existence of steady ridge
solutions will be discussed in the end of Sect. \ref{Sect. 5.3}.

\begin{figure}
\includegraphics[width=\columnwidth]{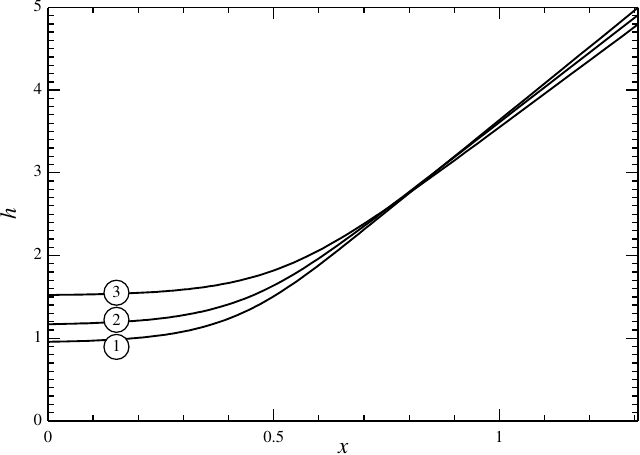}
\caption{The solution of the boundary-value problem (\ref{3.34})-(\ref{3.35}), (\ref{3.38}) for $\rho_{0}=0.85$ and (1) $\tau=0.05$, (2)
$\tau=0.1$, (3) $\tau=0.15$.}
\label{fig3}
\end{figure}

Eq. (\ref{3.34}) still admits solutions such that%
\begin{equation}
h\sim\theta x\qquad\text{as}\qquad x\rightarrow+\infty, \label{3.38}%
\end{equation}
where the constant $\theta$ can be identified with the contact angle (strictly
speaking, the contact angle equals $\arctan\theta$, but under the thin-film
approximation $\arctan\theta\approx\theta$). Examples of the solution of the
boundary-value problem (\ref{3.34}), (\ref{3.35}), (\ref{3.38}) has been
computed numerically and are shown in Fig. \ref{fig3}. Evidently, with
increasing temperature, the precursor film becomes thicker (see Fig.
\ref{fig3}a), whereas the contact angle becomes smaller (see Fig.
\ref{fig3}b). The latter conclusion agrees with the results of Ref.
\cite{Benilov20a} obtained for a realistic equation of state for water.

\section{Evolving interfaces: the regime with $\rho_{v}\sim\rho_{l}%
$\label{Sect. 4}}

Dynamics of liquid films depends strongly on the vapor-to-liquid density
ratio. The regime $\rho_{v}/\rho_{l}=\mathcal{O}(1)$ -- which occurs if
$\tau=\mathcal{O}(1)$ (i.e., the dimensional temperature $T$ is comparable to
the fluid's critical temperature $T_{cr}$) -- will be examined first. The
reader will see that, in this case, diffuse-interface films behave very
differently from their Navier--Stokes counterparts.

The asymptotic limit $T\ll T_{cr}$ will be examined in Sec. \ref{Sect. 5}.

\subsection{Nondimensionalization}

In addition to the nondimensional versions of coordinates (\ref{3.2}), density
(\ref{3.4}), and parameter $\rho_{0}$ (\ref{3.23}), introduce%
\begin{equation}
t_{nd}=\frac{\varepsilon^{2}v_{0}}{z_{0}}t,\qquad u_{nd}=\frac{u}%
{\varepsilon^{3}v_{0}},\qquad w_{nd}=\frac{w}{\varepsilon^{2}v_{0}%
},\label{4.1}%
\end{equation}%
\begin{equation}
T_{nd}=\frac{T}{T_{0}},\qquad p_{nd}=\frac{b^{2}}{a}p,\label{4.2}%
\end{equation}%
\begin{equation}
\left(  \mu_{s}\right)  _{nd}=\frac{\mu_{s}}{\mu_{0}},\qquad\left(  \mu
_{b}\right)  _{nd}=\frac{\mu_{b}}{\mu_{0}},\qquad\kappa_{nd}=\frac{\kappa
}{\kappa_{0}},\label{4.3}%
\end{equation}
where the shear and bulk viscosities are assumed to be of the same order
($\mu_{s}\sim\mu_{b}\sim\mu_{0}$) and%
\begin{equation}
v_{0}=\frac{az_{0}}{\mu_{0}b^{2}}\label{4.4}%
\end{equation}
is the general-case velocity scale (which applies when the film's slope
$\varepsilon$ and thickness $h$ are both order-one). Note that the powers of
$\varepsilon$ in (\ref{4.1}) have been chosen through the trial-and-error
approach, so that a consistent asymptotic model would be obtained in the end.

In terms of the nondimensional variables, the boundary-value problem
(\ref{2.1})-(\ref{2.8}) becomes (the subscript $_{nd}$ omitted)%
\begin{equation}
\frac{\partial\rho}{\partial t}+\fbox{$\varepsilon^2$}\frac{\partial\left(
\rho u\right)  }{\partial x}+\frac{\partial\left(  \rho w\right)  }{\partial
z}=0, \label{4.5}%
\end{equation}
\begin{widetext}%
\begin{multline}
\fbox{$\alpha\varepsilon^4$}\left(  \frac{\partial u}{\partial t}%
+\fbox{$\varepsilon^2$}u\frac{\partial u}{\partial x}+w\frac{\partial
u}{\partial z}\right)  +\frac{1}{\rho}\frac{\partial}{\partial x}\left(
\frac{\fbox{$\tau$}T\rho}{1-\rho}-\rho^{2}\right)  -\frac{\partial}{\partial
x}\left(  \fbox{$\varepsilon^2$}\frac{\partial^{2}\rho}{\partial x^{2}}%
+\frac{\partial^{2}\rho}{\partial z^{2}}\right)  \\
=\frac{\fbox{$\varepsilon^2$}}{\rho}\frac{\partial}{\partial x}\left[
2\fbox{$\varepsilon^2$}\mu_{s}\frac{\partial u}{\partial x}+\left(  \mu
_{b}-\frac{2}{3}\mu_{s}\right)  \left(  \fbox{$\varepsilon^2$}\frac{\partial
u}{\partial x}+\frac{\partial w}{\partial z}\right)  \right]  +\fbox{$%
\varepsilon^2$}\frac{\partial}{\partial z}\left[  \mu_{s}\left(
\frac{\partial u}{\partial z}+\frac{\partial w}{\partial x}\right)  \right]
,\label{4.6}%
\end{multline}%
\begin{multline}
\fbox{$\alpha\varepsilon^4$}\left(  \frac{\partial w}{\partial t}%
+\fbox{$\varepsilon^2$}u\frac{\partial w}{\partial x}+w\frac{\partial
w}{\partial z}\right)  +\frac{1}{\rho}\frac{\partial}{\partial z}\left(
\frac{\fbox{$\tau$}T\rho}{1-\rho}-\rho^{2}\right)  -\frac{\partial}{\partial
z}\left(  \fbox{$\varepsilon^2$}\frac{\partial^{2}\rho}{\partial x^{2}}%
+\frac{\partial^{2}\rho}{\partial z^{2}}\right)  \\
=\frac{\fbox{$\varepsilon^4$}}{\rho}\frac{\partial}{\partial x}\left[  \mu
_{s}\left(  \frac{\partial u}{\partial z}+\frac{\partial w}{\partial
x}\right)  \right]  +\fbox{$\varepsilon^2$}\frac{\partial}{\partial z}\left[
2\mu_{s}\frac{\partial w}{\partial z}+\left(  \mu_{b}-\frac{2}{3}\mu
_{s}\right)  \left(  \fbox{$\varepsilon^2$}\frac{\partial u}{\partial x}%
+\frac{\partial w}{\partial z}\right)  \right]  ,\label{4.7}%
\end{multline}%
\begin{multline}
\fbox{$\alpha\gamma\varepsilon^2$}\rho C_{V}\left(  \frac{\partial T}{\partial
t}+\fbox{$\varepsilon^2$}u\frac{\partial T}{\partial x}+w\frac{\partial
T}{\partial z}\right)  +\fbox{$\beta\varepsilon^2$}\frac{\fbox{$\tau$}T\rho
}{1-\rho}\left(  \fbox{$\varepsilon^2$}\frac{\partial u}{\partial x}%
+\frac{\partial w}{\partial z}\right)  \\
=\fbox{$\beta\varepsilon^4$}\left\{  \mu_{s}\left[  2\fbox{$\varepsilon
^4$}\left(  \frac{\partial u}{\partial x}\right)  ^{2}+\fbox{$\varepsilon
^2$}\left(  \frac{\partial u}{\partial z}+\frac{\partial w}{\partial
x}\right)  ^{2}+2\left(  \frac{\partial w}{\partial z}\right)  ^{2}\right]
+\left(  \mu_{b}-\frac{2}{3}\mu_{s}\right)  \left(  \fbox{$\varepsilon
^2$}\frac{\partial u}{\partial x}+\frac{\partial w}{\partial z}\right)
^{2}\right\}  \\
+\fbox{$\varepsilon^2$}\frac{\partial}{\partial x}\left(  \kappa\frac{\partial
T}{\partial x}\right)  +\frac{\partial}{\partial z}\left(  \kappa
\frac{\partial T}{\partial z}\right)  ,\label{4.8}%
\end{multline}
\end{widetext}%
\begin{equation}
u=0,\qquad w=0\qquad\text{at}\qquad z=0, \label{4.9}%
\end{equation}%
\begin{equation}
\rho=\rho_{0},\qquad T=1\qquad\text{at}\qquad z=0, \label{4.10}%
\end{equation}
where $\tau$ is given by (\ref{3.6}) and%
\begin{equation}
\alpha=\frac{K}{\mu_{0}^{2}b^{3}},\qquad\beta=\dfrac{aK}{\mu_{0}\kappa
_{0}T_{0}b^{4}}, \label{4.11}%
\end{equation}%
\begin{equation}
\gamma=\frac{c_{V}\mu_{0}}{\kappa_{0}},\qquad C_{V}=\frac{c_{V}}{R}.
\label{4.12}%
\end{equation}
Physically, $\alpha$ is the Reynolds number, $\gamma$ is the Prandtl number,
$C_{V}$ is the nondimensional heat capacity, and $\beta$ characterizes heat
release due to viscosity and fluid compression or cooling due to fluid expansion.

The parameter $\beta$ was first introduced in Refs.
\cite{Benilov20a,Benilov20b} for the case where the flow's aspect ratio was
order-one. It was concluded that $\beta$ is an `isothermality indicator': if
$\beta\sim1$, the effect of variable temperature is strong. The same is true
for liquid films -- despite the fact that Eq. (\ref{4.8}) and the boundary
condition (\ref{4.10}) suggest that the temperature is almost uniform -- i.e.,%
\begin{equation}
T=1+\varepsilon^{2}\tilde{T}(x,z,t).\label{4.13}%
\end{equation}
Yet, as seen below, the small variation $\tilde{T}$ affects the leading-order
film dynamics.

There is still a slight difference between liquid films and the general case:
in the latter, the heat production/consumption due to compressibility is
comparable to the heat production due to viscosity. In the liquid-film
equation (\ref{4.8}), on the other hand, the compressibility term exceeds the
viscosity term by an order of magnitude ($\varepsilon^{2}\beta$ to
$\varepsilon^{4}\beta$, respectively).

In what follows, $\beta$ is assumed to be order-one -- which it indeed is for
many common fluids (including water) at room temperature \cite{Benilov20b}. As
for $\alpha$ and $\gamma$, they appear in the governing equations only in a
product with a power of $\varepsilon$ -- so their values are unimportant as
long as they are not large (and, for common fluids, they are not
\cite{Benilov20b}). Finally, the nondimensional heat capacity $C_{V}$ will be
assumed to be order-one.

Another important feature of the proposed scaling is that the divergence terms
in the density equation (\ref{4.5}) are \emph{not} of the same order (as they
would be for Navier--Stokes films). This is due to the fact that, under the
regime considered, the interface is not driven by horizontal advection -- but
rather by evaporation and condensation, making it move vertically.\vspace
{0.5cm}

\subsection{The asymptotic equation\label{Sect. 4.2}}

Assume that the flow far above the substrate is not forced, so the viscous
stress is zero,%
\begin{equation}
\frac{\partial u}{\partial z}\rightarrow0,\qquad\frac{\partial w}{\partial
z}\rightarrow0\qquad\text{as}\qquad z\rightarrow\infty, \label{4.14}%
\end{equation}
and, as before, let%
\begin{equation}
\rho\rightarrow\rho_{v}\qquad\text{as}\qquad z\rightarrow\infty. \label{4.15}%
\end{equation}
In the study of static films in Sect. \ref{Sect. 3.2}, an `asymptotic
shortcut' has been used, and a similar one will be used for evolving films.

To derive it, multiply Eq. (\ref{4.7}) by $\left(  \rho-\rho_{0}\right)  $ and
integrate it from $z=0$ to $z=\infty$. Integrating the viscous term for $w$ by
parts and taking into account ansatz (\ref{4.13}) and the boundary conditions
(\ref{4.9})-(\ref{4.10}), (\ref{4.14})-(\ref{4.15}), one
obtains\begin{widetext}%
\begin{equation}
\varepsilon^{-2}\left\{  C-\frac{1}{2}\left[  \left(  \frac{\partial\rho
}{\partial z}\right)  _{z=0}\right]  ^{2}\right\}  +\frac{\partial}{\partial
x}\int_{0}^{\infty}\frac{\partial\rho}{\partial z}\frac{\partial\rho}{\partial
x}\mathrm{d}z+\tau\int_{0}^{\infty}\frac{\rho-\rho_{0}}{\rho}\frac{\partial
}{\partial z}\left(  \frac{\tilde{T}\rho}{1-\rho}\right)  \mathrm{d}z+\int%
_{0}^{\infty}\frac{\rho_{0}}{\rho^{2}}\frac{\partial\rho}{\partial z}%
\, \mu\frac{\partial w}{\partial z}\mathrm{d}z=\mathcal{O}(\varepsilon
^{2}),\label{4.16}%
\end{equation}
\end{widetext}where $C$ is given by (\ref{3.30}) and%
\[
\mu=\mu_{b}+\frac{4}{3}\mu_{s}.
\]
Next, observe that, to leading order, the dynamics equations (\ref{4.6}%
)-(\ref{4.7}) coincide with their static counterparts. As a result, the
density field of an evolving film is quasi-static and described by the
static-film expressions (\ref{3.27}) and (\ref{3.9}). The only difference is
that, the \emph{evolving}-film thickness $h$ should depend on $t$ as well as
$x$, so%
\begin{equation}
\rho=\bar{\rho}(z-h(x,t))+\mathcal{O}(\varepsilon). \label{4.17}%
\end{equation}
To obtain a closed-form equation for $h(x,t)$, it remains to express
$\tilde{T}$ and $w$ through $\rho$ and insert them into Eq. (\ref{4.16}).

To find $w$, substitute (\ref{4.17}) into Eq. (\ref{4.8}) and, taking into
account the boundary condition (\ref{4.9}), obtain%
\begin{equation}
w=\frac{\partial h}{\partial t}\frac{\bar{\rho}(z-h)-\rho_{0}}{\bar{\rho
}(z-h)}+\mathcal{O}(\varepsilon). \label{4.18}%
\end{equation}
Substitution of this expression and (\ref{4.17}) into Eq. (\ref{4.8}) yields%
\begin{multline}
\beta\tau\frac{\partial h}{\partial t}\frac{\rho_{0}\bar{\rho}(z-h)}%
{1-\bar{\rho}(z-h)}\frac{\partial\bar{\rho}(z-h)}{\partial z}\\
=\frac{\partial}{\partial z}\left[  \kappa(\bar{\rho}(z-h),1)\frac
{\partial\tilde{T}}{\partial z}\right]  +\mathcal{O}(\varepsilon),
\label{4.19}%
\end{multline}
where it has been taken into account that the dependence of the thermal
conductivity on the temperature is weak due to the near-isothermality
condition (\ref{4.13}).

One should assume that heat is neither coming from, nor going to, infinity,%
\[
\frac{\partial\tilde{T}}{\partial z}\rightarrow0\qquad\text{as}\qquad
z\rightarrow\infty,
\]
and also substitute (\ref{4.13}) into (\ref{4.10}) which yields%
\[
\tilde{T}=0\qquad\text{at}\qquad z=0.
\]
Solving Eq. (\ref{4.19}) with these boundary conditions, one obtains%
\begin{multline}
\tilde{T}=\beta\tau\frac{\partial h}{\partial t}\int_{0}^{z}\frac{\rho_{0}%
}{\kappa(\bar{\rho}(z_{1}-h),1)}\\
\times\left[  \ln\frac{\bar{\rho}(z_{1}-h)}{1-\bar{\rho}(z_{1}-h)}-\ln
\frac{\rho_{v}}{1-\rho_{v}}\right]  \mathrm{d}z_{1}+\mathcal{O}(\varepsilon).
\label{4.20}%
\end{multline}
Substituting expressions (\ref{4.18}) and (\ref{4.20}) into Eq. (\ref{4.16})
and keeping the leading-order terms only (which implies replacing $\rho_{0}$
with $\rho_{l}$), one obtains, after cumbersome but straightforward
algebra,\begin{widetext}%
\begin{multline}
\frac{\tau}{2\rho_{l}\left(  1-\rho_{l}\right)  ^{2}}-1-\frac{1}{2}%
\rho^{\prime2}-\frac{\partial^{2}h}{\partial x^{2}}\int_{0}^{\infty}\left[
\frac{\partial\bar{\rho}(z-h)}{\partial z}\right]  ^{2}\mathrm{d}%
z+\frac{\partial h}{\partial t}\int_{0}^{\infty}\frac{\rho_{l}^{2}\mu
(\bar{\rho}(z-h),1)}{\bar{\rho}^{4}(z-h)}\left[  \frac{\partial\bar{\rho
}(z-h)}{\partial z}\right]  ^{2}\mathrm{d}z\\
+\beta\tau^{2}\dfrac{\partial h}{\partial t}\int_{0}^{\infty}\frac{\rho_{l}%
}{\kappa(\bar{\rho}(z-h),1)}\left[  \ln\frac{\bar{\rho}(z-h)}{1-\bar{\rho
}(z-h)}-\ln\frac{\rho_{v}}{1-\rho_{v}}\right]  \left\{  \rho_{l}\left[
\ln\frac{\bar{\rho}(z-h)}{1-\bar{\rho}(z-h)}-\ln\frac{\rho_{v}}{1-\rho_{v}%
}\right]  -\frac{\rho_{l}-\rho_{v}}{1-\rho_{v}}\right\}  \mathrm{d}z\\
=0,\label{4.21}%
\end{multline}
\end{widetext}where $\rho^{\prime}(h)$ is the same function as its static-film
counterpart defined by (\ref{3.19}).

The integrals in this equality can be simplified using the assumption that $h$
exceeds the interfacial thickness. In the first two integrals, one can simply
move the lower limit to $-\infty$ and then replace $z-h$ with $z$ [the first
integral after that becomes equal to the surface tension $\sigma$ given by
(\ref{3.33})].

If, however, the same procedure is applied to the third integral in Eq.
(\ref{4.21}), it will diverge. To avoid the divergence and still take
advantage of $h$ being large, one should first use integration by parts (so
that the integrand is replaced by its derivative multiplied by $z$) and only
after that move the lower limit to $-\infty$. Eventually, one can transform
Eq. (\ref{4.21}) into\begin{widetext}%
\begin{equation}
\frac{\tau}{2\rho_{l}\left(  1-\rho_{l}\right)  ^{2}}-1-\frac{1}{2}%
\rho^{\prime2}(h)-\sigma\frac{\partial^{2}h}{\partial x^{2}}+\left[
A_{1}+\beta\left(  A_{2}+Bh\right)  \right]  \dfrac{\partial h}{\partial
t}=0,\label{4.22}%
\end{equation}
where%
\begin{equation}
A_{1}(\tau )=\int_{-\infty}^{\infty}\frac{\rho_{l}^{2}\mu(\bar{\rho}(z),1)}{\bar
{\rho}^{4}(z)}\left[  \frac{\mathrm{d}\bar{\rho}(z)}{\mathrm{d}z}\right]
^{2}\mathrm{d}z,\label{4.23}%
\end{equation}%
\begin{equation}
A_{2}(\tau )=-\tau^{2}\int_{-\infty}^{\infty}z\frac{\mathrm{d}}{\mathrm{d}z}\left(
\frac{\rho_{l}^{2}}{\kappa(\bar{\rho}(z),1)}\left\{  \frac{\bar{\rho
}(z)\left(  1-\rho_{v}\right)  }{\rho_{v}\left[  1-\bar{\rho}(z)\right]
}-\frac{\rho_{l}-\rho_{v}}{\rho_{l}\left(  1-\rho_{v}\right)  }\right\}
\ln\frac{\bar{\rho}(z)\left(  1-\rho_{v}\right)  }{\rho_{v}\left[  1-\bar
{\rho}(z)\right]  }\right)  \mathrm{d}z,\label{4.24}%
\end{equation}%
\begin{equation}
B(\tau )=\frac{\tau^{2}\rho_{l}^{2}}{\kappa(\rho_{l},1)}\left[  \ln\frac{\rho
_{l}\left(  1-\rho_{v}\right)  }{\rho_{v}\left(  1-\rho_{l}\right)  }%
-\frac{\rho_{l}-\rho_{v}}{\rho_{l}\left(  1-\rho_{v}\right)  }\right]
\ln\frac{\rho_{l}\left(  1-\rho_{v}\right)  }{\rho_{v}\left(  1-\rho
_{l}\right)  }.\label{4.25}%
\end{equation}
\end{widetext}Eq. (\ref{4.22}) is the desired asymptotic equation governing
$h(x,t)$; physically, it describes diffusion of chemical potential on its way
toward homogeneity.

\subsection{Discussion}

(1) Let us identify the physical meaning of the time-derivative term of Eq.
(\ref{4.22}) [the rest of the terms are the same as in the steady-state
equation (\ref{3.34})].

The term involving $A_{1}$ describes the interface's vertical motion driven by
evaporation and condensation, and the two terms multiplied by $\beta$ describe
heating/cooling of the fluid caused by its expansion/compression. Neither of
these effects is present in the Navier--Stokes films.\medskip

(2) Mathematically, Eq. (\ref{4.22}) is also very different from the equation
describing Navier--Stokes films. Even if the latter accounts for variable
temperature (as Eq. (1) of Ref. \cite{ThieleKnobloch04}), it does not involve
anything like the above-mentioned factor in front of $\partial h/\partial t$;
besides, it is of the fourth order in $x$, whereas Eq. (\ref{4.22}) is of the
second order. The dynamics described by the two models should be completely
different (this work is in progress).\medskip

\begin{figure*}
\includegraphics[width=\textwidth]{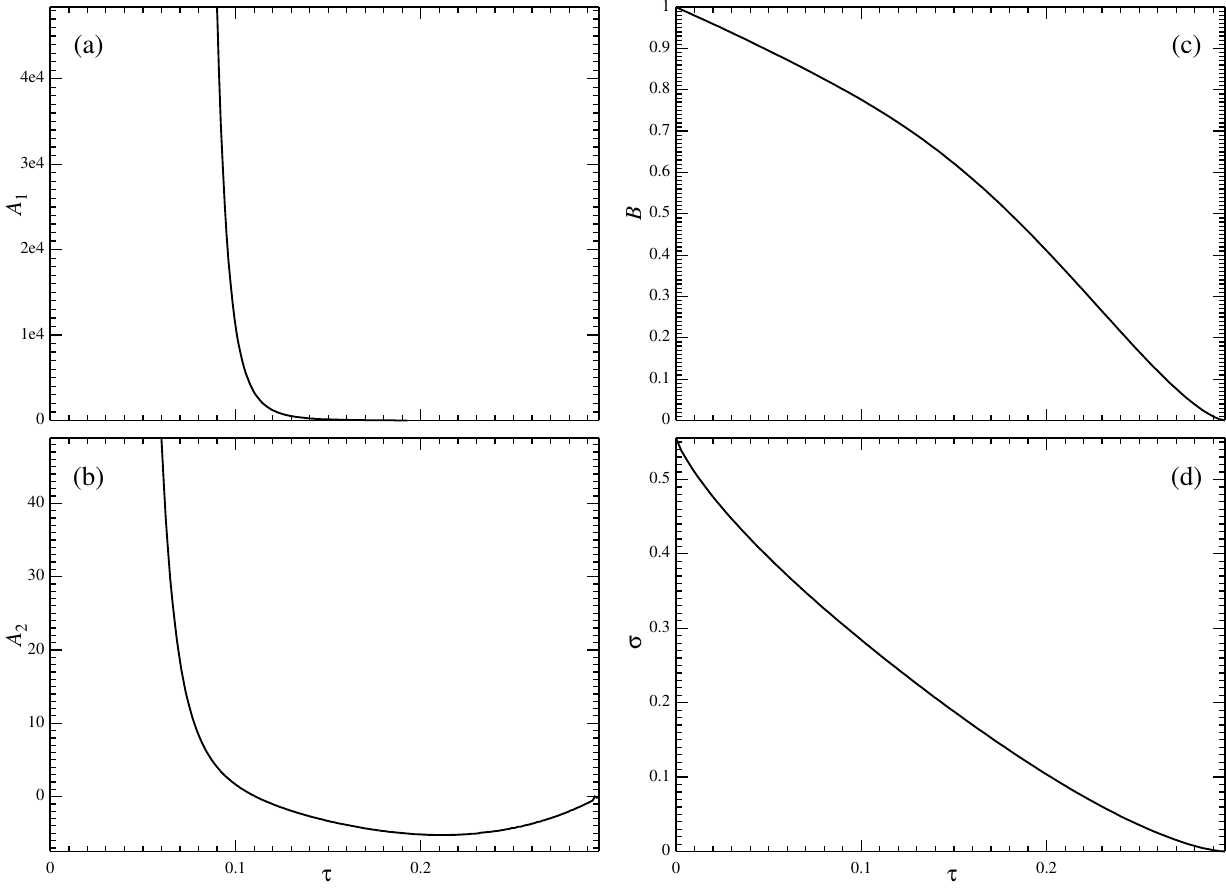}
\caption{The coefficients of Eq. (\ref{4.22}) [with $\mu$ and $\kappa$ given by (\ref{4.26})] vs. the nondimensional temperature $\tau$. (a) $A_{1}$ given by (\ref{4.23}), (b) $A_{2}$ given by (\ref{4.24}), (c) $B$
given by (\ref{4.25}), (d) $\sigma$ (the surface tension) given by (\ref{3.32}).}
\label{fig4}
\end{figure*}

(3) It is instructive to compute the coefficients of Eq. (\ref{4.22}). To do
so, one has to specify the effective viscosity $\mu$ and thermal conductivity
$\kappa$ -- for example, assume that they are proportional to the fluid
density. The proportionality coefficients should generally depend on the
temperature, but due to the near-isothermality ansatz (\ref{4.13}) the
temperature is close to being constant -- hence, can be eliminated by a proper
choice of the nondimensionalization scales $\mu_{0}$ and $\kappa_{0}$. Thus,
one can simply let%
\begin{equation}
\mu(\rho,1)=\rho,\qquad\kappa(\rho,1)=\rho. \label{4.26}%
\end{equation}
The coefficients $A_{1}$, $A_{2}$, $B$, and $\sigma$ -- given by
(\ref{4.23})-(\ref{4.25}) and (\ref{3.33}), respectively -- have been computed
and are plotted in Fig. \ref{fig4}. Observe that, as $\tau\rightarrow0$, the
coefficient $A_{1}$ grows, as does $A_{2}$ (although much slower than $A_{1}$)
-- whereas $B$ and $\sigma$ remain finite. The limits of the latter two can be
calculated using the small-$\tau$ asymptotics (\ref{3.15})-(\ref{3.17}), which
yield%
\begin{equation}
B\rightarrow1,\qquad\sigma\rightarrow2^{-5/2}\pi\qquad\text{as}\qquad
\tau\rightarrow0. \label{4.27}%
\end{equation}
The reason why $A_{1}$ and $A_{2}$ are singular as $\tau\rightarrow0$ can be
readily seen from expressions (\ref{4.23}) and (\ref{4.24}), which both
involve division by $\bar{\rho}(z)$ -- whose minimum value, $\rho_{v}$, tends
to zero as $\tau\rightarrow0$. In fact, one can derive asymptotically (see
Appendix \ref{Appendix C}) that%
\begin{equation}
A_{1}\approx0.36994\,\tau^{1/2}\rho_{_{v}}^{-3/2}\qquad\text{if}\qquad\tau
\ll1. \label{4.28}%
\end{equation}%
\begin{equation}
A_{2}\approx0.75918\,\tau^{3/2}\rho_{v}^{-1/2}\qquad\text{if}\qquad\tau\ll1.
\label{4.29}%
\end{equation}
The singular behavior of $A_{1}$ and $A_{2}$ indicates that Eq. (\ref{4.22})
fails when the temperature is low enough to make $\rho_{v}$ small; in terms of
the dimensional variables, Eq. (\ref{4.22}) fails when the vapor-to-liquid
density ratio is small. What happens in this case is examined in the next section.

\section{Regime(s) with $\rho_{v}\ll\rho_{l}$\label{Sect. 5}}

It turns out that the asymptotic regime corresponding to the limit%
\begin{equation}
\rho_{v}\rightarrow0,\qquad\varepsilon=\operatorname{const}\label{5.1}%
\end{equation}
does not `overlap' with the limit%
\[
\rho_{v}=\operatorname{const},\qquad\varepsilon\rightarrow0
\]
examined previously. This suggests that there exists an intermediate regime,
where $\rho_{v}$ is small, but is still comparable to, say, a certain power of
$\varepsilon$.

It is worth noting that, in all three regimes, the leading-order solution is
represented by $\bar{\rho}(z-h)$ where $\bar{\rho}(z)$ decribes a liquid/vapor
interface in an unbounded space. The difference in the value of $\tau$,
however, makes $\bar{\rho}(z)$ specific to the corresponding regime -- which,
in turn, affects higher orders.

In what follows, limit (\ref{5.1}) will be examined in Sects. \ref{Sect. 5.1}%
-\ref{Sect. 5.3}, whereas the intermediate regime will be examined in Sects.
\ref{Sect. 5.4}-\ref{Sect. 5.5}.

\subsection{Regime (\ref{5.1}): the nondimensionalization\label{Sect. 5.1}}

As seen earlier, smallness of $\tau$ implies \emph{exponential} smallness of
$\rho_{v}$ -- thus, when they appear in the same expression, the former should
be by comparison treated as an order-one quantity. Another important point is
that the smallness of $\rho_{v}$ does not affect the scaling of $\rho(x,z,t)$
whose maximum value remains to be order-one. In fact, only the velocity and
time need to be rescaled -- by switching to the same scaling as that for the
Navier--Stokes films.

Summarizing the above, one should revise the finite-$\rho_{v}$ scaling by
replacing (\ref{4.1}) with%
\begin{equation}
t_{nd}=\frac{\varepsilon^{4}v_{0}}{z_{0}}t,\qquad u_{nd}=\frac{u}%
{\varepsilon^{3}v_{0}},\qquad w_{nd}=\frac{w}{\varepsilon^{4}v_{0}}.
\label{5.2}%
\end{equation}
The resulting nondimensional equations are%
\begin{equation}
\frac{\partial\rho}{\partial t}+\frac{\partial\left(  \rho u\right)
}{\partial x}+\frac{\partial\left(  \rho w\right)  }{\partial z}=0,
\label{5.3}%
\end{equation}
\begin{widetext}%
\begin{multline}
\fbox{$\alpha\varepsilon^6$}\left(  \frac{\partial u}{\partial t}%
+u\frac{\partial u}{\partial x}+w\frac{\partial u}{\partial z}\right)
+\frac{1}{\rho}\frac{\partial}{\partial x}\left(  \frac{\tau T\rho}{1-\rho
}-\rho^{2}\right)  -\frac{\partial}{\partial x}\left(  \fbox{$\varepsilon
^2$}\frac{\partial^{2}\rho}{\partial x^{2}}+\frac{\partial^{2}\rho}{\partial
z^{2}}\right)  \\
=\frac{\fbox{$\varepsilon^4$}}{\rho}\frac{\partial}{\partial x}\left[
2\mu_{s}\frac{\partial u}{\partial x}+\left(  \mu_{b}-\frac{2}{3}\mu
_{s}\right)  \left(  \frac{\partial u}{\partial x}+\frac{\partial w}{\partial
z}\right)  \right]  +\frac{\fbox{$\varepsilon^2$}}{\rho}\frac{\partial
}{\partial z}\left[  \mu_{s}\left(  \frac{\partial u}{\partial z}%
+\fbox{$\varepsilon^2$}\frac{\partial w}{\partial x}\right)  \right]
,\label{5.4}%
\end{multline}%
\begin{multline}
\fbox{$\alpha\varepsilon^8$}\left(  \frac{\partial w}{\partial t}%
+u\frac{\partial w}{\partial x}+w\frac{\partial w}{\partial z}\right)
+\frac{1}{\rho}\frac{\partial}{\partial z}\left(  \frac{\tau T\rho}{1-\rho
}-\rho^{2}\right)  -\frac{\partial}{\partial z}\left(  \fbox{$\varepsilon
^2$}\frac{\partial^{2}\rho}{\partial x^{2}}+\frac{\partial^{2}\rho}{\partial
z^{2}}\right)  \\
=\frac{\fbox{$\varepsilon^4$}}{\rho}\left\{  \frac{\partial}{\partial
x}\left[  \mu_{s}\left(  \frac{\partial u}{\partial z}+\fbox{$\varepsilon
^2$}\frac{\partial w}{\partial x}\right)  \right]  +\frac{\partial}{\partial
z}\left[  2\mu_{s}\frac{\partial w}{\partial z}+\left(  \mu_{b}-\frac{2}{3}%
\mu_{s}\right)  \left(  \frac{\partial u}{\partial x}+\frac{\partial
w}{\partial z}\right)  \right]  \right\}  ,\label{5.5}%
\end{multline}%
\begin{multline}
\fbox{$\alpha\gamma\varepsilon^4$}\rho C_{V}\left(  \frac{\partial T}{\partial
t}+u\frac{\partial T}{\partial x}+w\frac{\partial T}{\partial z}\right)
+\fbox{$\beta\varepsilon^4$}\frac{\tau T\rho}{1-\rho}\left(  \frac{\partial
u}{\partial x}+\frac{\partial w}{\partial z}\right)  \\
=\fbox{$\beta\varepsilon^6$}\left\{  \mu_{s}\left[  2\fbox{$\varepsilon
^2$}\left(  \frac{\partial u}{\partial x}\right)  ^{2}+\left(  \frac{\partial
u}{\partial z}+\fbox{$\varepsilon^2$}\frac{\partial w}{\partial x}\right)
^{2}+2\fbox{$\varepsilon^2$}\left(  \frac{\partial w}{\partial z}\right)
^{2}\right]  +\fbox{$\varepsilon^2$}\left(  \mu_{b}-\frac{2}{3}\mu_{s}\right)
\left(  \frac{\partial u}{\partial x}+\frac{\partial w}{\partial z}\right)
^{2}\right\}  \\
+\fbox{$\varepsilon^2$}\frac{\partial}{\partial x}\left(  \kappa\frac{\partial
T}{\partial x}\right)  +\frac{\partial}{\partial z}\left(  \kappa
\frac{\partial T}{\partial z}\right)  ,\label{5.6}%
\end{multline}
\end{widetext}where the parameters $\tau$, $\alpha$, $\beta$, $\gamma$, and
$C_{V}$ are determined by (\ref{3.6}) and (\ref{4.11})-(\ref{4.12}). The
boundary conditions look the same as those for the finite-$\rho_{v}$ regime --
see (\ref{4.9})-(\ref{4.10}) and (\ref{4.14})-(\ref{4.15}).

\subsection{Regime (\ref{5.1}): the asymptotic equation\label{Sect. 5.2}}

Eq. (\ref{5.6}) suggests that%
\begin{equation}
T=1+\varepsilon^{4}\tilde{T}(x,z,t),\label{5.7}%
\end{equation}
Comparison of ansatz (\ref{5.7}) and its the finite-$\rho_{v}$ analogue
(\ref{4.13}) shows that the temperature variations are now weaker than those
in the finite-$\rho_{v}$ regime.

Substituting (\ref{5.7}) into Eqs. (\ref{5.4})-(\ref{5.5}), one can rewrite
them in the form%
\begin{multline}
\frac{\partial}{\partial x}\left[  \tau\left(  \ln\frac{\rho}{1-\rho}+\frac
{1}{1-\rho}\right)  -2\rho-\varepsilon^{2}\frac{\partial^{2}\rho}{\partial
x^{2}}-\frac{\partial^{2}\rho}{\partial z^{2}}\right] \\
=\frac{\varepsilon^{2}}{\rho}\frac{\partial}{\partial z}\left[  \mu_{s}%
(\rho,1)\frac{\partial u}{\partial z}\right]  +\mathcal{O}(\varepsilon^{4}),
\label{5.8}%
\end{multline}%
\begin{multline}
\frac{\partial}{\partial z}\left[  \tau\left(  \ln\frac{\rho}{1-\rho}+\frac
{1}{1-\rho}\right)  -2\rho-\varepsilon^{2}\frac{\partial^{2}\rho}{\partial
x^{2}}-\frac{\partial^{2}\rho}{\partial z^{2}}\right] \\
=\mathcal{O}(\varepsilon^{4}), \label{5.9}%
\end{multline}
Observe that $\tilde{T}$ does not appear in the leading and next-to-leading
orders of these equation, with the implication that the non-isothermality
effect is now too weak to affect interfacial dynamics.

It follows from Eq. (\ref{5.9}) that, to leading order, the expression in the
square brackets is a function of $x$ and $t$ (but not $z$), with Eq.
(\ref{5.8}) suggesting that this function is $\mathcal{O}(\varepsilon^{2})$.
Thus, denoting it by $\varepsilon^{2}F(x,t)$, one can rewrite Eqs.
(\ref{5.8})-(\ref{5.9}) in the form%
\begin{equation}
\varepsilon^{2}\frac{\partial F(x,t)}{\partial x}=\frac{\varepsilon^{2}}{\rho
}\frac{\partial}{\partial z}\left[  \mu_{s}(\rho,1)\frac{\partial u}{\partial
z}\right]  +\mathcal{O}(\varepsilon^{4}), \label{5.10}%
\end{equation}%
\begin{multline}
\tau\left(  \ln\frac{\rho}{1-\rho}+\frac{1}{1-\rho}\right)  -2\rho
-\varepsilon^{2}\frac{\partial^{2}\rho}{\partial x^{2}}-\frac{\partial^{2}%
\rho}{\partial z^{2}}\\
=\varepsilon^{2}F(x,t)+\mathcal{O}(\varepsilon^{4}). \label{5.11}%
\end{multline}
Assuming as before that $h\gg1$, one can replace $\rho$ with $\bar{\rho}(z-h)$
and then use Eq. (\ref{5.11}) to relate $F$ to $h$. To do so, multiply
(\ref{5.11}) by $\partial\rho/\partial z$, integrate with respect to $z$ from
$0$ to $\infty$, and use the boundary conditions, which yields%
\begin{multline}
F(x,t)=\frac{1}{\rho_{0}-\rho_{v}}\left[  \sigma\frac{\partial^{2}h}{\partial
x^{2}}+\frac{1}{2}\rho_{0}^{\prime2}(h)\right] \\
+\operatorname{const}+\mathcal{O}(\varepsilon^{2}), \label{5.12}%
\end{multline}
where the surface tension $\sigma$ is given by (\ref{3.33}) and the specific
expression for $\operatorname{const}$ will not be needed.

Under the same assumption $h\gg1$, one can let $\rho=\bar{\rho}%
(z-h)+\mathcal{O}(\varepsilon)$ and $\bar{\rho}(-h)=\rho_{l}+\mathcal{O}%
(\varepsilon)$. Keeping in mind these equalities, one can use Eqs.
(\ref{5.11}), (\ref{5.3}), and the boundary conditions to
deduce\begin{widetext}%
\begin{equation}
u=-\frac{\partial F}{\partial x}\int_{0}^{z}\frac{1}{\mu_{s}(\bar{\rho}%
(z_{2}-h),1)}\int_{z_{2}}^{\infty}\left[  \bar{\rho}(z_{1}-h)-\rho_{v}\right]
\mathrm{d}z_{1}\mathrm{d}z_{2}+\mathcal{O}(\varepsilon),\label{5.13}%
\end{equation}%
\begin{equation}
w=\frac{1}{\bar{\rho}(z-h)}\left\{  \left[  \rho_{l}-\bar{\rho}(z-h)\right]
\frac{\partial h}{\partial t}+\frac{\partial}{\partial x}\int_{0}^{z}\bar
{\rho}(z_{1}-h)\,u(x,z_{1},t)\,\mathrm{d}z_{1}\right\}  +\mathcal{O}%
(\varepsilon).\label{5.14}%
\end{equation}
Substituting the former expression into the latter and introducing an
auxiliary function%
\begin{equation}
\bar{\varrho}(z)=\int_{z}^{\infty}\left[  \bar{\rho}(z_{1})-\rho_{v}\right]
\mathrm{d}z_{1},\label{5.15}%
\end{equation}
one can obtain (after straightforward algebra)%
\begin{equation}
w=-\frac{1}{\bar{\rho}(z-h)}\left\{  \left[  \bar{\rho}(z-h)-\rho_{l}\right]
\frac{\partial h}{\partial t}+\frac{\partial}{\partial x}\left[
\frac{\partial F}{\partial x}\int_{0}^{z}\frac{\bar{\varrho}(z-h)-\bar
{\varrho}(z_{1}-h)}{\mu_{s}(\bar{\rho}(z_{1}-h),1)}\bar{\varrho}%
(z_{1}-h)\,\mathrm{d}z_{1}\right]  \right\}  +\mathcal{O}(\varepsilon
).\label{5.16}%
\end{equation}
\end{widetext}One can now take advantage of the assumption $\tau\ll1$ and,
thus, replace $\bar{\rho}$ with asymptotic (\ref{3.17}). Among other things,
it implies that $\bar{\rho}(z-h)\rightarrow0$ as $z\rightarrow h+2^{-3/2}\pi$
-- which gives rise to a singularity in expression (\ref{5.16}). To avoid the
singularity, one has to assume
\begin{multline}
\rho_{l}\frac{\partial h}{\partial t}+\frac{\partial}{\partial x}\left[
\frac{\partial F}{\partial x}\int_{0}^{h+2^{-3/2}\pi}\frac{\bar{\varrho}%
^{2}(z_{1}-h)}{\mu_{s}(\bar{\rho}(z_{1}-h),1)}\mathrm{d}z_{1}\right] \\
=0. \label{5.17}%
\end{multline}
To simplify this equation, observe that expressions (\ref{3.15}) and
(\ref{3.21}) imply%
\[
\rho_{l}=1+\mathcal{O}(\tau),\qquad\rho_{0}=1+\mathcal{O}(\tau,\varepsilon).
\]
Now, replacing in Eq. (\ref{5.17}) $F$ and $\varrho$ with expressions
(\ref{5.12}) and (\ref{5.15}), respectively, and keeping the leading-order
terms only, one obtains%
\begin{equation}
\frac{\partial h}{\partial t}+\frac{\partial}{\partial x}\left[
Q(h)\frac{\partial}{\partial x}\left(  \sigma\frac{\partial^{2}h}{\partial
x^{2}}+\frac{1}{2}\rho^{\prime2}\right)  \right]  =0, \label{5.18}%
\end{equation}
where%
\begin{equation}
Q(h)=\int_{-h}^{2^{-3/2}\pi}\left[  \int_{z}^{2^{-3/2}\pi}\bar{\rho}%
(z_{1})\,\mathrm{d}z_{1}\right]  ^{2}\frac{\mathrm{d}z}{\mu_{s}(\bar{\rho
}(z),1)}, \label{5.19}%
\end{equation}
and $\bar{\rho}(z)$ is given by (\ref{3.17}). Eq. (\ref{5.18}) is the desired
asymptotic equation for $h(x,t)$.

It is instructive to calculate the function $Q(h)$ for a particular case --
say,%
\[
\mu_{s}(\rho,1)=q\rho,
\]
where $q$ is a constant. Then, expression (\ref{5.19}) yields%
\begin{align}
Q(h)  &  =q^{-1}\left[  \frac{1}{3}h^{3}+2^{-1/2}\pi\left(  1-\ln2\right)
-2^{-9/2}3^{-1}\pi^{3}\right] \nonumber\\
&  \approx q^{-1}\left(  \frac{1}{3}h^{3}+0.22489\right)  . \label{5.20}%
\end{align}
Even though this expression was derived under the assumption that $h$ is
large, $h$ may be \emph{logarithmically} large -- hence, the retainment of the
constant in the above expression is justified. For the same reason one may
want to keep in Eq. (\ref{5.18}) $\sigma$ instead of replacing it with its
small-$\tau$ limit (\ref{4.27}).

\subsection{Regime (\ref{5.1}): existence of liquid ridges\label{Sect. 5.3}}

Steady-state solutions of Eq. (\ref{5.18}) satisfy%
\begin{equation}
\sigma\frac{\mathrm{d}^{2}h}{\mathrm{d}x^{2}}+\frac{1}{2}\rho^{\prime2}=D,
\label{5.21}%
\end{equation}
where $h=h(x)$ and $D>0$ is a constant of integration. The mere fact that Eq.
(\ref{5.21}) involves an arbitrary constant [unlike its finite-$\rho_{v}$
counterpart (\ref{3.34})] allows the ridge solution to exist. It can be
readily shown that, if%
\[
0<\left(  2D\right)  ^{1/2}<\max\left\{  \rho^{\prime}(h)\right\}  ,
\]
(\ref{5.21}) admits a symmetric solution such that%
\[
h\rightarrow h_{pf}\qquad\text{as}\qquad x\rightarrow\pm\infty,
\]
where $h_{pf}$ is the smaller root of the equation%
\[
\rho^{\prime2}(h_{pf})=2D.
\]
In addition to $h_{pf}$, this equation has another (larger) root -- say,
$h_{i}$. Recalling Fig. \ref{fig2} (which shows what the graph of the function
$\rho^{\prime}(h)$ looks like), one can see $\rho^{\prime}(h_{pf})<0$, whereas
$\rho^{\prime}(h_{i})>0$. Obviously, $h_{i}$ corresponds to the inflection
point of $h(x)$.

The ridge solution can be found in an implicit form by reducing (\ref{5.21})
to a first-order separable equation.

Thus, the asymptotic model for the case $\rho_{v}/\rho_{l}\ll1$ admits steady
solutions -- whereas the $\rho_{v}/\rho_{l}\sim1$ model (examined in Sect.
\ref{Sect. 3.3}) does not. This suggests that, in the exact equations, the
ridges exist as \emph{quasi}-steady solutions: generally, they evolve (so are
\emph{not} steady) -- but, if $\rho_{v}/\rho_{l}\ll1$, their evolution is slow
and indistinguishable from, say, evaporation. Given that, for a drop of water
on one's kitchen table, $\rho_{v}/\rho_{l}$ is indeed small, this argument
should help to reconcile the unexpected mathematical results obtained in this
paper and Ref. \cite{Benilov20c} with one's everyday intuition.

\subsection{The intermediate regime: the asymptotic equation\label{Sect. 5.4}}

Note that the small-$\rho_{v}$ equation (\ref{5.18}) cannot be obtained from
its finite-$\rho_{v}$ counterpart (\ref{4.22}) by letting $\rho_{v}%
\rightarrow0$. This suggest that there may exist an intermediate regime.

Finding this regime is not straightforward, however. Firstly, there are three
small parameters in the problem: $\varepsilon$, $\rho_{v}$, and $1/h$, making
a formal expansion cumbersome even if $\tau$ [related to $\rho_{v}$ through
equality (\ref{3.16})] is treated as an order-one parameter. Secondly, the
regions where $\rho\sim1$ and $\rho\ll1$ are to be examined differently,
implying a convoluted matching procedure.

To find a reasonably simple approach to exploring the intermediate regime,
recall that the finite- and small-$\rho_{v}$ limits differ by the scaling of
the vertical velocity $w$ [compare (\ref{4.1}) and (\ref{5.2})]. Thus, the
intermediate regime can be found by considering the small-$\rho_{v}$ equations
(\ref{5.3})-(\ref{5.6}), but retaining the terms involving $w$ even if they
appear to be of a higher-order in $\varepsilon$.

Accordingly, rewrite Eqs. (\ref{5.4})-(\ref{5.5}) in the
form\begin{widetext}%
\begin{multline}
\frac{\partial}{\partial x}\left[  \tau\left(  \ln\frac{\rho}{1-\rho}+\frac
{1}{1-\rho}\right)  -2\rho-\varepsilon^{2}\frac{\partial^{2}\rho}{\partial
x^{2}}-\frac{\partial^{2}\rho}{\partial z^{2}}\right]  =\frac{\varepsilon^{4}%
}{\rho}\frac{\partial}{\partial x}\left[  \left(  \mu_{b}-\frac{2}{3}\mu
_{s}\right)  \frac{\partial w}{\partial z}\right]  \\
+\frac{\varepsilon^{2}}{\rho}\frac{\partial}{\partial z}\left[  \mu_{s}%
(\rho,1)\left(  \frac{\partial u}{\partial z}+\varepsilon^{2}\frac{\partial
w}{\partial x}\right)  \right]  +\mathcal{O}(\varepsilon^{4},\varepsilon
^{6}w),\label{5.22}%
\end{multline}%
\begin{equation}
\frac{\partial}{\partial z}\left[  \tau\left(  \ln\frac{\rho}{1-\rho}+\frac
{1}{1-\rho}\right)  -2\rho-\varepsilon^{2}\frac{\partial^{2}\rho}{\partial
x^{2}}-\frac{\partial^{2}\rho}{\partial z^{2}}\right]  =\frac{\varepsilon^{4}%
}{\rho}\frac{\partial}{\partial z}\left[  \mu(\rho,1)\frac{\partial
w}{\partial z}\right]  +\mathcal{O}(\varepsilon^{4},\varepsilon^{6}%
w),\label{5.23}%
\end{equation}
where, as before, $\mu=\mu_{b}+4\mu_{s}/3$. Eq. (\ref{5.23}) can be used to
derive the `asymptotic shortcut': multiplying (\ref{5.23}) by $\left(
\rho-\rho_{0}\right)  $ and carrying out straightforward algebra [similar to
that in Sects. \ref{Sect. 3.2} and \ref{Sect. 4.2})], one obtains
\begin{equation}
\frac{\tau}{2\rho_{l}\left(  1-\rho_{l}\right)  ^{2}}-1-\frac{1}{2}%
\rho^{\prime2}-\sigma\frac{\partial^{2}h}{\partial x^{2}}+\varepsilon^{2}%
\int_{0}^{\infty}\frac{\rho_{l}}{\bar{\rho}^{2}(z-h)}\frac{\partial\bar{\rho
}(z-h)}{\partial z}\mu(\bar{\rho}(z-h),1)\frac{\partial w}{\partial
z}\mathrm{d}z=\mathcal{O}(\varepsilon^{2},\varepsilon^{4}w).\label{5.24}%
\end{equation}
\end{widetext}To reduce Eq. (\ref{5.24}) to a closed-form equation for $h$,
one should first use Eq. (\ref{5.3}) to relate $w$ to $\bar{\rho}(z-h)$ and
$u$, and then use Eq. (\ref{5.22}) to relate $u$ to $\bar{\rho}(z-h)$.
Unfortunately, the latter equation -- unlike its small-$\rho_{v}$ counterpart
(\ref{5.8}) -- includes $w$, making it impossible to eliminated it after all.

Luckily, the contribution of $w$ to Eq. (\ref{5.22}) turns out to be negligible.

To understand why, recall that expression (\ref{5.14}) for $w$ applies to both
previously-considered limits -- hence, it applies to the intermediate regime
also. Using it and the leading-order solution (\ref{3.17}) for $\bar{\rho}$,
one obtains%
\[
w=\mathcal{O}(\rho_{v}^{-1})\qquad\text{at}\qquad z\approx h+2^{-3/2}\pi,
\]%
\[
w=\mathcal{O}(1)\qquad\text{at}\qquad z\not \approx h+2^{-3/2}\pi.
\]
Thus, $w$ has a peak near $z=h+2^{-3/2}\pi$, and it can be further estimated
(see Appendix \ref{Appendix D}) that the characteristic width of this peak is
$\tau^{-1/2}\rho_{v}^{1/2}$.

Next, the expression in the square brackets in Eq. (\ref{5.22}) can be denoted
(as before) by $\varepsilon^{2}F(x,t)$. Considering the resulting equation as
a means of finding $u$, one can see that it involves two components:

\begin{enumerate}
\item a contribution of the term involving $F$ (this component is of order-one
and is spread between $z=0$ and $z\approx h+2^{-3/2}\pi$), and

\item a contribution of the term involving $w$ (of amplitude $\varepsilon
^{2}\rho_{v}^{-1}$ and width $\tau^{-1/2}\rho_{v}^{1/2}$ localized near the
point $z=h+2^{-3/2}\pi$).
\end{enumerate}

Once $u$ is substituted into Eq. (\ref{5.14}), both components are multiplied
by $\bar{\rho}(z-h)$ and integrated -- thus, component 1 contributes
$\mathcal{O}(1)$, whereas component 2 contributes $\mathcal{O}(\varepsilon
^{2}\rho_{v}^{1/2}\tau^{-1/2})$. The latter is smaller -- which effectively
means that the $w$-involving terms in Eq. (\ref{5.22}) can be omitted -- which
effectively means that, in the intermediate regime, $w$ can still be
approximated by the small-$\rho_{v}$ expression (\ref{5.16}).

Substituting (\ref{5.16}) into Eq. (\ref{5.24}), one
obtains\begin{widetext}%
\begin{equation}
\frac{\tau}{2\rho_{l}\left(  1-\rho_{l}\right)  ^{2}}-1-\frac{1}{2}%
\rho^{\prime2}-\sigma\frac{\partial^{2}h}{\partial x^{2}}+\varepsilon^{2}%
A_{1}\left\{  \frac{\partial h}{\partial t}+\frac{\partial}{\partial x}\left[
Q(h)\frac{\partial}{\partial x}\left(  \sigma\frac{\partial^{2}h}{\partial
x^{2}}+\frac{1}{2}\rho^{\prime2}\right)  \right]  \right\}  =0,\label{5.25}%
\end{equation}
\end{widetext}where $A_{1}$ is given by expression (\ref{4.23}) and $Q(h)$, by
(\ref{5.19}).

Eq. (\ref{5.25}) is the asymptotic equation governing the intermediate regime.

\subsection{The intermediate regime: discussion\label{Sect. 5.5}}

(1) In principle, $\rho_{l}$, $\sigma$, and $A_{1}$ in Eq. (\ref{5.25}) can be
replaced with their small-$\tau$ estimates (\ref{3.15}), (\ref{4.27}), and
(\ref{4.28}), respectively. Using the last of the three estimates and, for
simplicity, treating $\tau$ as an order-one parameter, one can see that the
last term in (\ref{5.25}) is order-one when%
\begin{equation}
\rho_{_{v}}\sim\varepsilon^{4/3}. \label{5.26}%
\end{equation}
This is the applicability condition of the intermediate regime examined in
this section, whereas the finite-$\rho_{v}$ regime and the small-$\rho_{v}$
limit are valid if $\rho_{v}\gg\varepsilon^{4/3}$ and $\rho_{v}\ll
\varepsilon^{4/3}$, respectively.\medskip

(2) Note that Eq. (\ref{5.25}) was obtained under the assumption that the
near-isothermality ansatz (\ref{5.7}) used for the small-$\rho_{v}$ regime
applies to the intermediate regime as well. This can be verified through an
asymptotic analysis of the temperature equation (\ref{5.6}), in a manner
similar to how Eq. (\ref{5.22}) was analyzed.\medskip

(3) It is unlikely that Eq. (\ref{5.25}) admits solutions describing liquid
ridges, but their nonexistence it is not easy to prove.

\section{Three-dimensional liquid films\label{Sect. 6}}

Even though the asymptotic equations (\ref{4.22}), (\ref{5.18}), and
(\ref{5.25}) have been derived for two-dimensional films, they can be readily
extended to three dimensions. In what follows, these 3D extensions are summarized.

In the main body of the paper, two of the asymptotic equations derived are
written in nondimensional variables that are different from those of the third
equation. In this section, all equations are written in terms of the variables
for the small-$\rho_{v}$ regime [i.e., those given by (\ref{3.2})-(\ref{3.3}),
(\ref{3.23}), (\ref{5.2}), and (\ref{4.2})-(\ref{4.4})].

The nondimensional parameter space of the problem involves the vapor-to-liquid
density ratio, $\rho_{v}/\rho_{l}$, and the parameter $\varepsilon$ defined by
(\ref{3.21}) (physically, the latter is proportional to the contact angle).
Since this paper deals with \emph{thin} liquid films, $\varepsilon\ll1$.

The limit $\rho_{v}/\rho_{l}\gg\varepsilon^{4/3}$ was examined in Sect.
\ref{Sect. 4}, and the 3D extension of the asymptotic equation (\ref{4.22})
derived there is%
\begin{multline}
\varepsilon^{2}\left[  A_{1}+\beta\left(  A_{2}+Bh\right)  \right]
\dfrac{\partial h}{\partial t}\\
-\sigma\nabla^{2}h+\frac{\tau}{2\rho_{l}\left(  1-\rho_{l}\right)  ^{2}%
}-1-\frac{1}{2}\rho^{\prime2}(h)=0, \label{6.1}%
\end{multline}
where the nondimensional temperature $\tau$ is given by (\ref{3.6}), the
surface tension $\sigma$ and the coefficients $A_{1}$, $A_{2}$, and $B$ depend
on $\tau$ and are given by expressions (\ref{3.33}) and (\ref{4.23}%
)-(\ref{4.25}), respectively. The function $\rho^{\prime}(h)$ (see examples in
Fig. \ref{fig2}) is defined by (\ref{3.19}) and determined by the
boundary-value problem (\ref{3.10})-(\ref{3.13}).

The regime $\rho_{v}/\rho_{l}\sim\varepsilon^{4/3}$ was examined in Sects.
\ref{Sect. 5.4}-\ref{Sect. 5.5}. The 3D extension of the asymptotic equation
(\ref{5.25}) is%
\begin{multline}
\varepsilon^{2}A_{1}\left(  \frac{\partial h}{\partial t}+\mathbf{\nabla}%
\cdot\left\{  Q(h)\mathbf{\nabla}\left[  \sigma\nabla^{2}h+\frac{1}{2}%
\rho^{\prime2}(h)\right]  \right\}  \right) \\
-\sigma\nabla^{2}h+\frac{\tau}{2\rho_{l}\left(  1-\rho_{l}\right)  ^{2}%
}-1-\frac{1}{2}\rho^{\prime2}(h)=0, \label{6.2}%
\end{multline}
where the function $Q(h)$ is determined by (\ref{5.19}).

Finally, the limit $\rho_{v}/\rho_{l}\ll\varepsilon^{4/3}$ was examined in
Sects. \ref{Sect. 5.1}-\ref{Sect. 5.3}, and the 3D extension of the asymptotic
equation (\ref{5.18}) is%
\begin{equation}
\frac{\partial h}{\partial t}+\mathbf{\nabla}\cdot\left\{  Q(h)\mathbf{\nabla
}\left[  \sigma\nabla^{2}h+\frac{1}{2}\rho^{\prime2}(h)\right]  \right\}  =0.
\label{6.3}%
\end{equation}
For macroscopic films -- such that the film thickness exceeds that of the
liquid/vapor interface by several order of magnitude -- one can assume in
expression (\ref{5.19}) that $\bar{\rho}\approx\rho_{l}$ and thus obtain%
\[
Q(h)\approx\frac{\rho_{l}^{2}}{3\mu_{s}}h^{3},
\]
where $\mu_{s}$ is the nondimensional shear viscosity of the liquid phase.
Furthermore, in the limit $h\rightarrow\infty$, the function $\rho^{\prime
}(h)$ tends to a constant (see Fig. \ref{fig2}). As a result, Eq. (\ref{6.3})
reduces to the equation for the usual Navier--Stokes films,%
\[
\frac{\partial h}{\partial t}+\frac{\rho_{l}^{2}\sigma}{3\mu_{s}%
}\mathbf{\nabla}\cdot\left(  h\mathbf{\nabla}\nabla^{2}h\right)  =0.
\]
This conclusion helps to understand why the Navier--Stokes equations follow
from the DIM in the incompressibility limit, but -- unlike the DIM -- admit
solutions describing liquid ridges.

Indeed, for common fluids at room temperature, $\rho_{v}/\rho_{l}$ is very
small: for water at $T=20^{\circ}\mathrm{C}$, for example, $\rho_{v}/\rho
_{l}\approx1.7\times10^{-5}$. An estimate of $\varepsilon$, in turn, can be
deduced from the fact that contact angles of common fluids on commonly used
substrates are unlikely to be smaller than $5^{\circ}$. This implies that
liquid ridges can be modelled using Eq. (\ref{6.2}) with (sic!) $\varepsilon
^{2}A_{1}\gg1$. Consequently, the terms in Eq. (\ref{6.2}) that prevent liquid
ridges from being steady are small and the resulting evolution is slow --
probably indistinguishable from evaporation and other effects not taken into
account by the present model.

\section{Concluding remarks}

Thus, three parameter regimes have been identified and three asymptotic models
have been derived for liquid films. Two points are still in order: one on the
results obtained and another, on how to improve them.

(1) One should realize that the diffuse-interface model (used to derive all of
the results of the present work) does not include any adjustable parameters,
i.e., such that could be used to optimize the results to fit a specific
phenomenon. In addition to the equation of state (typically, known from
thermodynamics handbooks), the DIM includes only the Korteweg parameter $K$
and the near-wall density $\rho_{0}$. The former is uniquely linked to the
surface tension of the fluid under consideration and the latter, to the static
contact angle.

(2) Before applying the present results to a specific fluid, one should make
them more realistic -- by extending them to a mixture of several fluids and
assume the temperature to be subcritical for one fluid and supercritical for
all the others. Such a model should provide a sufficient accurate description
of, say, a water droplet surrounded by air, at a room temperature.

\section{Data availability}

The data that support the findings of this study are available from the corresponding author upon reasonable request.

\appendix

\section{The Maxwell construction\label{Appendix A}}

It follows from (\ref{3.10})-(\ref{3.12}) that%
\begin{multline}
\tau\left(  \ln\frac{\rho_{v}}{1-\rho_{v}}+\frac{1}{1-\rho_{v}}\right)
-2\rho_{v}\\
=\tau\left(  \ln\frac{\rho_{l}}{1-\rho_{l}}+\frac{1}{1-\rho_{l}}\right)
-2\rho_{l}. \label{A.1}%
\end{multline}
Physically, Eq. (\ref{A.1}) means that the free-energy density of the vapor
phase equals that of the liquid.

Another equation inter-relating $\rho_{v}$ and $\rho_{l}$ can be obtained by
considering%
\[
\int_{-\infty}^{\infty}\bar{\rho}\times\frac{\mathrm{d}}{\mathrm{d}z}\left(
\ref{3.10}\right)  \mathrm{d}z.
\]
Integrating the term involving $\mathrm{d}^{2}\bar{\rho}/\mathrm{d}z^{2}$ by
parts and taking into account the boundary conditions (\ref{3.11}%
)-(\ref{3.12}), one obtains%
\begin{equation}
\frac{\tau\rho_{v}}{1-\rho_{v}}-\rho_{v}^{2}=\frac{\tau\rho_{l}}{1-\rho_{l}%
}-\rho. \label{A.2}%
\end{equation}
Physically, (\ref{A.2}) is the condition of equality of the pressure in the
vapor phase to that in the liquid phase. In this paper, Eqs. (\ref{A.1}%
)-(\ref{A.2}) are referred to as the Maxwell construction.

In the low-temperature limit, one expects%
\begin{equation}
\rho_{v}\rightarrow0,\qquad\rho_{l}\rightarrow1\qquad\tau\rightarrow0.
\label{A.3}%
\end{equation}
Under the assumption that, as $\tau\rightarrow\infty$, $\rho_{v}$ becomes
exponentially small (to be verified later), Eq. (\ref{A.2}) yields%
\begin{equation}
\rho_{l}=\frac{1+\sqrt{1-4\tau}}{2}+\mathcal{O}(\tau\rho_{v}). \label{A.4}%
\end{equation}
Next, rearranging Eq. (\ref{A.1}) using (\ref{A.3}), one obtains%
\begin{equation}
\rho_{v}=\frac{\rho_{l}}{1-\rho_{l}}\exp\left[  -\frac{2\rho_{l}}{\tau}%
+\frac{\rho_{l}}{1-\rho_{l}}+\mathcal{O}(\rho_{v})\right]  . \label{A.5}%
\end{equation}
Using the leading-order term of (\ref{A.4}) to rearrange the leading-order
term of (\ref{A.5}) and using the leading-order term of the latter to
rearrange the error in both expressions, one obtains (\ref{3.15})-(\ref{3.16})
as required.

\section{Properties of $\rho(z|h)$\label{Appendix B}}

The function $\rho(z|h)$ is determined by the boundary-value problem
(\ref{3.5}), (\ref{2.8}), (\ref{3.7})-(\ref{3.8}) -- which can actually be
solved analytically, albeit in an implicit form. To do so, multiply
(\ref{3.5}) by $\mathrm{d}\rho/\mathrm{d}z$, integrate with respect to $z$
and, recalling condition (\ref{2.8}) and definition (\ref{3.19}) of
$\rho^{\prime}(h)$ obtain%
\begin{equation}
\left(  \frac{\mathrm{d}\rho}{\mathrm{d}z}\right)  ^{2}=F(\rho), \label{B.1}%
\end{equation}
where%
\begin{multline}
F(\rho)=\left(  \rho_{l}-\rho_{0}\right)  ^{2}\rho^{\prime2}+2\left[  \rho
_{0}^{2}-\rho^{2}-G\left(  \rho-\rho_{0}\right)
\vphantom{\left(\frac{\rho}{1-\rho}\right) }\right. \\
+\left.  \tau\left(  \rho\ln\frac{\rho}{1-\rho}-\rho_{0}\ln\frac{\rho_{0}%
}{1-\rho_{0}}\right)  \right]  . \label{B.2}%
\end{multline}
Eq. (\ref{B.1}) is separable (hence, can be solved analytically), but it
involves an unknown constant $G$. To determine it, introduce%
\begin{equation}
\rho_{_{\infty}}=\lim\limits_{z\rightarrow\infty}\rho(z|h) \label{B.3}%
\end{equation}
(note that, generally, $\rho_{\infty}\neq\rho_{v}$), and observe that Eq.
(\ref{3.5}) implies that%
\begin{equation}
G=\tau\left(  \ln\frac{\rho_{\infty}}{1-\rho_{\infty}}+\frac{1}{1-\rho
_{\infty}}\right)  -2\rho_{_{\infty}}. \label{B.4}%
\end{equation}
Note also that (\ref{B.3}) is consistent with Eq. (\ref{B.1}) only if
$F(\rho_{\infty})=0$ -- which yields [together with expressions (\ref{B.2})
and (\ref{B.4})]\begin{widetext}%
\begin{equation}
\left(  \rho_{l}-\rho_{0}\right)  ^{2}\rho^{\prime2}+2\left(  \rho_{0}%
-\rho_{\infty}\right)  ^{2}+2\tau\left[  \rho_{0}\ln\dfrac{\rho_{\infty
}\left(  1-\rho_{0}\right)  }{\rho_{0}\left(  1-\rho_{\infty}\right)  }%
+\frac{\rho_{0}-\rho_{\infty}}{1-\rho_{\infty}}\right]  =0.\label{B.5}%
\end{equation}\end{widetext}This equation relates $\rho_{\infty}$ to $\rho
_{0}$ and $\rho^{\prime}$. If $\rho^{\prime}<0$, $\rho(z|h)$ monotonically
decays with $z$ -- but, if $\rho^{\prime}>0$, $\rho(z|h)$ has a maximum.
Common sense and numerical experiments suggest that, when increasing
$\rho^{\prime}$, this maximum approaches $\rho_{l}$, whereas $\rho_{\infty}$
approaches $\rho_{v}$ (while the distance $h$ between the substrate and
interface tends to infinity). Thus, denoting the upper bound of $\rho^{\prime
}$ by $\rho_{max}^{\prime}$, one can find it by substituting $\rho^{\prime
}=\rho_{max}^{\prime}$ and $\rho_{\infty}=\rho_{v}$ into (\ref{B.5}). Finally,
expressing $\rho_{max}^{\prime}$ from the resulting equation, one can obtain
estimate (\ref{3.20}) as required.

\section{Estimates (\ref{4.28})-(\ref{4.29})\label{Appendix C}}

In a manner similar to how Eqs. (\ref{B.1})-(\ref{B.2}) were obtained, one can
use Eq. (\ref{3.10}) and the boundary condition (\ref{3.12}) to obtain%
\[
\frac{\mathrm{d}\bar{\rho}}{\mathrm{d}z}=-2^{1/2}\left\{  \tau\left[
\bar{\rho}\ln\frac{\bar{\rho}\left(  1-\rho_{v}\right)  }{\rho_{v}\left(
1-\bar{\rho}\right)  }-\frac{\bar{\rho}-\rho_{_{v}}}{1-\rho_{v}}\right]
-\left(  \bar{\rho}-\rho_{_{v}}\right)  ^{2}\right\}  ^{1/2}.
\]
Using this equality and omitting overbars, one can rewrite (\ref{4.23}%
)-(\ref{4.24}) and (\ref{4.26}) in the form\begin{widetext}%
\begin{equation}
A_{1}=2^{1/2}\rho_{l}^{2}\int_{\rho_{v}}^{\rho_{l}}\frac{1}{\rho^{3}\left(
1-\rho\right)  }\left\{  \tau\left[  \rho\ln\frac{\rho\left(  1-\rho
_{v}\right)  }{\rho_{v}\left(  1-\rho\right)  }-\frac{\rho-\rho_{_{v}}}%
{1-\rho_{v}}\right]  -\left(  \rho-\rho_{_{v}}\right)  ^{2}\right\}
^{1/2}\mathrm{d}\rho,\label{C.1}%
\end{equation}%
\begin{equation}
A_{2}=\tau^{2}\rho_{l}^{2}\int_{\rho_{l}}^{\rho_{v}}\frac{z(\rho)}{\rho^{2}%
}\left\{  \left[  \ln\frac{\rho\left(  1-\rho_{v}\right)  }{\rho_{v}\left(
1-\rho\right)  }-\frac{\rho_{l}-\rho_{v}}{\rho_{l}\left(  1-\rho_{v}\right)
}-\frac{2}{1-\rho}\right]  \ln\frac{\rho\left(  1-\rho_{v}\right)  }{\rho
_{v}\left(  1-\rho\right)  }+\frac{\rho_{l}-\rho_{v}}{\rho_{l}\left(
1-\rho_{v}\right)  \left(  1-\rho\right)  }\right\}  \mathrm{d}\rho
.\label{C.2}%
\end{equation}
\end{widetext}Observe that, in (\ref{C.2}), $z$ is a function of $\rho$.

In the limit $\tau\rightarrow0$, the main contribution in integrals
(\ref{C.1})-(\ref{C.2}) comes from the neighborhood of the point $\rho
=\rho_{v}$, which suggests the substitution: $\rho=\rho_{_{v}}\xi$. Keeping in
(\ref{C.1})-(\ref{C.2}) the leading order only, one obtains%
\begin{equation}
A_{1}=\frac{2^{1/2}\tau^{1/2}}{\rho_{_{v}}^{3/2}}\int_{1}^{\infty}\frac{1}%
{\xi^{3}}\left[  \xi\left(  \ln\xi-1\right)  +1\right]  ^{1/2}\mathrm{d}\xi,
\label{C.3}%
\end{equation}%
\begin{equation}
A_{2}=-\frac{\tau^{2}}{\rho_{v}}\int_{1}^{\infty}\frac{z(\rho_{_{v}}\xi)}%
{\xi^{2}}\left[  \left(  \ln\xi-3\right)  \ln\xi+1\right]  \mathrm{d}\xi,
\label{C.4}%
\end{equation}
Evaluating the integral in (\ref{C.4}) numerically, one obtains (\ref{4.28}).

To derive (\ref{4.29}), observe that it follows from the linearized version of
Eq. (\ref{3.10}) that%
\begin{multline*}
\bar{\rho}\sim\rho_{v}\\
+\Delta\exp\left[  -\sqrt{\frac{\tau}{\rho_{v}\left(  1-\rho_{v}\right)  ^{2}%
}-2}z\right]  \qquad\text{as}\qquad z\rightarrow\infty,
\end{multline*}
whereas monotonicity of $\bar{\rho}(z)$ implies that the constant $\Delta$ is
positive. Letting $\bar{\rho}=\rho_{_{v}}\xi$ and taking into account that
$\rho_{v}\ll1$, one obtains%
\begin{equation}
z\sim-\rho_{v}^{1/2}\tau^{-1/2}\ln\left(  \xi-1\right)  -\ln\Delta.
\label{C.5}%
\end{equation}
Substituting (\ref{C.5}) into (\ref{C.4}), one can verify that the integral
involving $D$ vanishes, and%
\[
A_{2}=\rho_{v}^{-1/2}\tau^{3/2}\int_{1}^{\infty}\frac{\ln\left(  \xi-1\right)
}{\xi^{2}}\left[  \left(  \ln\xi-3\right)  \ln\xi+1\right]  \mathrm{d}\xi.
\]
Finally, evaluating the integral in the above expression numerically, one
obtains (\ref{4.29}), as required.

\section{The asymptotics of $\bar{\rho}(z)$ as $\bar{\rho}\rightarrow\rho_{v}%
$\label{Appendix D}}

The asymptotics of the peak of $w$ [given by expression (\ref{5.14})] is
determined by the region where $\bar{\rho}(z)$ is small. To examine it,
consider Eq. (\ref{3.10}) for $\bar{\rho}(z)$ and let $\bar{\rho}=\rho
_{v}\tilde{\rho}$. Then, taking into account (\ref{3.26}) and keeping the
leading-order terms only, one obtains%
\[
\ln\tilde{\rho}+\frac{\rho_{v}}{\tau}\frac{\mathrm{d}^{2}\tilde{\rho}%
}{\mathrm{d}z^{2}}=0.
\]
Evidently, all small parameters can be scaled out from this equation by
changing $z$ to $\xi$ such that%
\[
\xi=\left(  \frac{\tau}{\rho_{v}}\right)  ^{1/2}\left(  z-2^{-3/2}\right)  .
\]
This effectively means that the characteristic width of the small-$\bar{\rho}$
region is $\left(  \rho_{v}/\tau\right)  ^{1/2}$.

\bibliography{.././../bib/refs}

\end{document}